\def\attunearxiv{}
\documentclass{article}
\usepackage[T1]{fontenc}
\usepackage[utf8]{inputenc}
\usepackage{microtype}
\usepackage{graphicx}
\usepackage{booktabs}
\usepackage{array}
\usepackage{tabularx}
\usepackage{amsmath}
\usepackage{amssymb}
\usepackage{enumitem}
\usepackage{xcolor}
\usepackage{hyperref}
\usepackage{xurl}
\usepackage{needspace}

\ifdefined\attunearxiv
  \usepackage[accepted]{mlsys2025}
  \hypersetup{pdfsubject={}}%
  \makeatletter
  \renewcommand{\Notice@String}{Preprint.}%
  \makeatother
\else
  \usepackage{mlsys2025}
\fi
\mlsystitlerunning{ATTUNE: Agentic Throughput Tuning for LLM Training}

\makeatletter
\newcommand{\attunepos}[4]{}
\newcommand{\attunegeom}[4]{}
\newcommand{\attunemark}[1]{\leavevmode\pdfsavepos
  \write\@auxout{\string\attunepos{#1}{\thepage}{\the\pdflastxpos}{\the\pdflastypos}}}
\AtBeginDocument{\protected@write\@auxout{}{\string\attunegeom{\the\dimexpr 1in+\voffset+\topmargin+\headheight+\headsep\relax}{\the\textheight}{\the\paperheight}{\the\paperwidth}}}
\makeatother

\begin{document}
\makeatletter%
\setlength{\@fptop}{0pt}\setlength{\@fpsep}{14pt plus 4pt}\setlength{\@fpbot}{0pt plus 1fil}
\setlength{\@dblfptop}{0pt}\setlength{\@dblfpsep}{14pt plus 4pt}\setlength{\@dblfpbot}{0pt plus 1fil}
\makeatother

\twocolumn[
\mlsystitle{ATTUNE: Agentic Throughput Tuning for LLM Training}

\makeatletter
\ifdefined\isaccepted
  \renewcommand{\@pa}[1]{}
  \gdef\printAffiliationsAndNotice#1{{\let\thefootnote\relax\footnotetext{\hspace*{-\footnotesep}%
    Correspondence to: \mlsyscorrespondingauthor@text.

    \ \\
    \Notice@String}}}
\fi
\makeatother
\begin{mlsysauthorlist}
\mlsysauthor{Leo Fan}{amzn}
\mlsysauthor{Hongyang Wang}{amzn}
\mlsysauthor{Wenbo Chen}{amzn}
\mlsysauthor{Jingguo Liu}{amzn}
\mlsysauthor{Xian Li}{amzn}
\mlsysauthor{Xuan Wang}{amzn}
\mlsysauthor{Ruochen Wen}{amzn}
\mlsysauthor{Yiming Yan}{amzn}
\mlsysauthor{Hao Wu}{amzn}
\mlsysauthor{Yichuan Li}{amzn}
\mlsysauthor{Qian Qu}{amzn}
\mlsysauthor{Zheshen Wang}{amzn}
\mlsysauthor{Benoit Dumoulin}{amzn}
\ifdefined\isaccepted\\[2pt] Amazon\fi
\end{mlsysauthorlist}

\ifdefined\isaccepted\else\mlsysaffiliation{amzn}{Amazon}\fi

\mlsyscorrespondingauthor{Leo Fan\newline}{\mbox{jianhfan@amazon.com}, \mbox{leofanxyz@gmail.com}}

\mlsyskeywords{LLM training, throughput tuning, coding agents, Machine Learning and Systems}

\vskip 0.3in

\begin{abstract}
Training a large language model can occupy thousands of GPUs for weeks, and its speed depends on its recipe: how the model is split across GPUs, how memory is traded for compute, and which kernels and communication overlaps it uses. Finding a good recipe is still manual work for experts. Automated tuners search fixed lists of knobs and check only that a job runs, so they miss large gains and can ship changes that silently alter what the model learns. ATTUNE tunes any working recipe unattended within a GPU-hour budget. A coding agent, guided by lessons from past tuning runs, finds each run's bottleneck and changes one setting at a time, and a correctness gate outside the agent refuses changes known to alter training. ATTUNE is deployed in production at Amazon Ads, a business that reported \$19.8 billion in revenue in Q2 2026. Its recipes train the models that power Amazon's Ads Agent. On three of the models it tunes, its recipes raise throughput per GPU by 1.7$\times$ on average and up to 3.9$\times$ over internal expert-tuned recipes. It also speeds up four of NVIDIA's benchmark recipes by 7--26\% and held-out public recipes by up to 2.6$\times$, and beats Bayesian optimization, random search and OptiCo, a recent LLM-agent tuner, at the same GPU-hour budget. We release ATTUNE at \href{https://github.com/amazon-science/attune}{\texttt{\mbox{github.com/}\allowbreak{}\mbox{amazon-science/}\allowbreak{}\mbox{attune}}} and OpenRecipes, a leaderboard of more than 60 verified recipes, at \href{https://openrecipes.ai}{\texttt{\mbox{openrecipes.ai}}}.
\end{abstract}
]

\printAffiliationsAndNotice{}

\section{Introduction}\label{sec:1}

\begin{figure}[!t]
\centering
\includegraphics[width=\columnwidth]{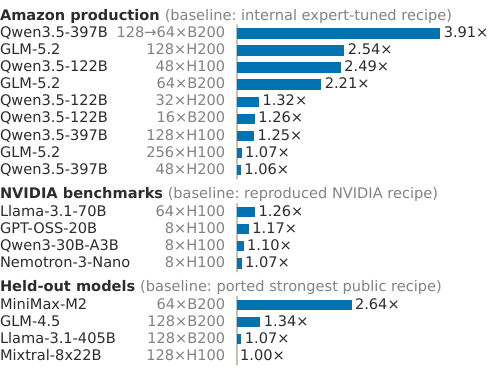}
\caption{ATTUNE speeds up the expert-tuned recipes of nine Amazon production jobs by 1.7$\times$ on average, four of NVIDIA's benchmark recipes by 7--26\%, and four held-out public recipes by up to 2.6$\times$. Each bar is the shipped recipe's tokens/s/GPU over its starting recipe's. The tuned Qwen3.5-397B recipe on B200 runs on half the GPUs. On Mixtral-8x22B, no candidate beat the public recipe by more than run-to-run noise. See Section~\ref{sec:3.6} and Table~\ref{tab:1}.}\label{fig:gains}
\end{figure}

As large language models (LLMs) have grown to hundreds of billions of parameters, their training jobs occupy hundreds to thousands of GPUs, and the efficiency of each job has become a first-order cost concern. A job's throughput is set by its training \emph{recipe}, the configuration that jointly chooses the tensor, pipeline, expert, context and data parallelism degrees, the microbatch size, which activations are recomputed or offloaded, which kernels run, and how communication overlaps with compute. Each choice trades memory against compute or communication, and the best mix depends on the model, the GPUs and the workload. In practice, engineers find a good recipe by trial and error: they launch multi-node runs, profile them, look for memory limits or communication stalls, adjust the configuration and repeat, often over days of cluster time. The resulting recipes are rarely published and seldom carry over to another model or cluster, so most teams repeat this work for their own jobs.

Existing tools automate recipe tuning, but each addresses only part of the problem. Parallelization planners \citep{zheng2022alpa,miao2022galvatron,liu2024aceso,um2024metis} search parallel layouts with a cost model, a simulator or per-stage profiling, but they cover little beyond placement and do not search kernels or other software-stack settings. Vendor autoconfigurators \citep{deepspeed2021deepspeed,nvidia2026nemo} measure candidates on the cluster, but only over a small grid of documented knobs such as parallel degrees and microbatch size. Black-box search such as Bayesian optimization \citep{bergstra2011algorithms} needs no model of the system but learns only from scores, so it spends many costly runs, many of which run out of memory. Recent tuners built on LLM agents, OptiCo \citep{chen2026optico}, PROMPTS \citep{ding2026prompts} and ASAP \citep{ding2025asap}, do reason from run measurements, but they tune only a fixed list of knobs and check at most that a configuration runs, not that it still trains the same model.

Two observations from tuning Amazon's production recipes motivate a different design. First, large gains often come from settings these lists leave out, such as splitting each pipeline stage into smaller virtual stages or switching the library that sends tokens to experts. The 3.9$\times$ production gain in Figure~\ref{fig:gains}, for example, includes three changes that OptiCo's 11 knobs cannot make: virtual stages, turning CPU offload off and halving the GPUs. Second, a faster configuration can silently change what the model learns. An expert capacity factor, a speed option for mixture-of-experts (MoE) models, drops the tokens that overflow an expert, and the flag \texttt{-{}-grad-\allowbreak{}reduce-\allowbreak{}in-\allowbreak{}bf16}, which looks like a communication setting, also turns off FP32 gradient accumulation. Neither change makes a job fail, so a check that the job runs would ship both.

We present ATTUNE, an agentic tuner that takes a working training recipe, including one written by an expert, and makes it faster on the user's own workload and software stack. Each tuning campaign runs unattended from one command to a shipped recipe within a GPU-hour budget the user sets. A single general-purpose coding agent runs the search, guided by a tuning \emph{skill}: instructions and scripts with the search rules and a catalog that records what each \emph{lever}, a setting the search can change, did in past campaigns. After each run, the agent finds the bottleneck and changes one lever at a time. Because it reads the framework source, including the code behind each flag before using it, it can change any argument the framework exposes.

\begin{figure*}[!t]
\centering
\includegraphics[width=\textwidth]{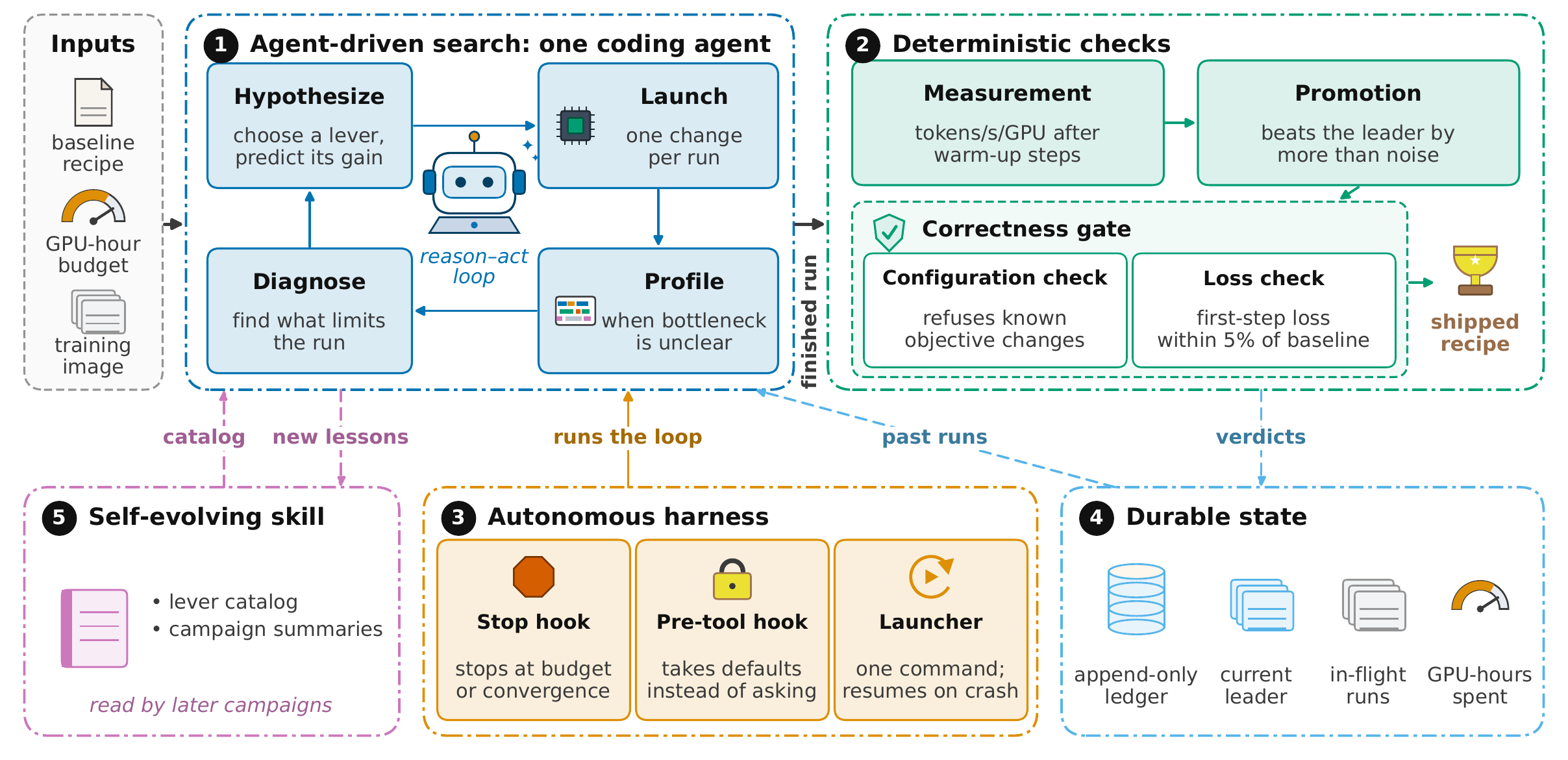}
\caption{ATTUNE's architecture. (1) A coding agent runs the search: it states a hypothesis, launches one change, profiles the run when the bottleneck is unclear and diagnoses it. (2) Deterministic checks outside the agent refuse objective-changing flags, decide which measurements count and which candidate leads, and check each candidate before it can ship. (3) The harness keeps the loop running on its own, (4) durable state lets a campaign resume, and (5) the skill carries the lever catalog and past campaigns' lessons to later campaigns. The leader is the best candidate so far.}\label{fig:1}
\end{figure*}

A correctness gate checks that every recipe ATTUNE ships still trains the same objective. Before launch, a configuration check refuses any argument known to change the objective or to trigger a known defect in the stack. After the run, a loss check rejects the recipe if its loss at the first training step differs from the starting recipe's by more than 5\%.

We evaluate ATTUNE in Amazon's production post-training jobs and on two sets of public recipes (Figure~\ref{fig:gains}). The first set is four of NVIDIA's NeMo AutoModel benchmark recipes, run unchanged on NVIDIA's framework and the H100s they were written for. The second holds four models that ATTUNE was not developed on, MiniMax-M2 \citep{minimax2025minimaxm2}, GLM-4.5 \citep{glm452025glm45}, Llama-3.1-405B \citep{llama2024llama} and Mixtral-8x22B \citep{jiang2024mixtral,mistral2024cheaper}, each with its strongest public recipe ported to our stack and trained on real text. On MiniMax-M2 and GLM-4.5, we also compare ATTUNE with OptiCo, Bayesian optimization and random search.

Our contributions are:

\begin{itemize}
\item \textbf{ATTUNE}, an agentic tuner that takes any working training recipe to a faster one unattended. Its coding agent, guided by a tuning skill, can change any framework argument, and a correctness gate outside the agent refuses changes known to alter training.
\item \textbf{Production deployment at Amazon.} ATTUNE is deployed in Amazon's production post-training, where its recipes train models that power Amazon's Ads Agent \citep{amazon2026ads}. On three of the models it tunes, its recipes are 1.7$\times$ faster on average. We share lessons from running it.
\item \textbf{An extensive evaluation and an open leaderboard.} ATTUNE speeds up NVIDIA's benchmark recipes by 7--26\% and held-out public recipes by up to 2.6$\times$, and beats Bayesian optimization, random search and OptiCo under the same GPU-hour budget. We release ATTUNE and OpenRecipes, a leaderboard of more than 60 verified recipes, so that teams can start from a tuned recipe instead of repeating the search.
\end{itemize}

\section{ATTUNE Design}\label{sec:2}

ATTUNE separates recipe tuning into an agent-driven search and deterministic checks. A coding agent, guided by a tuning skill, runs the search: it diagnoses each run, forms a hypothesis, changes the recipe and launches the job. Deterministic code decides which measurements are valid, which candidate leads and which recipe ships. A harness keeps the loop running without supervision (Figure~\ref{fig:1}).

A \emph{campaign} tunes one model on one cluster, starting from a working \emph{baseline} recipe and running under a GPU-hour budget. Each \emph{candidate} recipe runs a short benchmark, and the best recipe so far is the \emph{leader}, the one each new candidate must beat. The campaign ends when the budget is spent, or earlier if the agent judges that the search has converged.

\subsection{Search}\label{sec:2.1}

\subsubsection{Search Loop}\label{sec:2.1.1}

The agent works in a reason--act loop: it reasons about the latest result, then acts with ordinary tools, namely a shell, the cluster API, and file search and reading. The file tools give it the framework source inside the training image, so it can read how any flag is implemented. The tuning skill supplies the search procedure, the measurement rules, the lever catalog and helper scripts. The skill instructs the agent to prioritize candidates by expected gain and confidence, estimated from four sources: its diagnosis of what limits the leader, the lever catalog, the ledger of the campaign's past runs and the flags in the framework source. For each lever, the catalog records how it works, when it helps, and the direction and rough size of its effect.

\paragraph{Search discipline.} The skill gives the agent six rules:

\begin{enumerate}
\item State a hypothesis before launching each candidate: the bottleneck it should shrink, the mechanism and the expected gain.
\item Change one lever, or one coupled set of parallelism degrees, per candidate, so each result can be attributed.
\item Before using a flag, read its implementation in the running image, including what it asserts and which flags it silently overrides.
\item Rule out a lever only after measuring two of its values on the same recipe, or after reaching the lever's feasible limit. Conclude that the layout is exhausted only after probing each of tensor, pipeline, expert, context and data parallelism (TP, PP, EP, CP, DP; \citealp{shoeybi2019megatronlm,narayanan2021efficient,rajbhandari2022deepspeedmoe,liu2025moe}).
\item When trying a new layout, first turn off the leader's memory-saving settings (e.g., full recomputation), unless a memory estimate shows the layout needs them. If it then runs out of memory, turn them back on and measure again before ruling it out.
\item Once the layout stops changing, re-test the flags that won on earlier layouts, since a flag's gain often does not carry over to another layout.
\end{enumerate}

\subsubsection{Diagnosis and Profiling}\label{sec:2.1.2}

The agent diagnoses each run from its metrics and training log, mainly step time, per-stage memory and step-time variance across the measured steps. When the bottleneck is still unclear, it profiles a run, tracing one rank per pipeline stage. A deterministic trace analyzer splits each traced stage's step into compute, communication, host-side and idle time, so the agent sees which stage is slowest and what it waits on.

\subsection{Measurement and Control Plane}\label{sec:2.2}

\paragraph{Workload.} A campaign keeps its workload fixed across candidates: the model, dataset, sequence length, global batch and precision. Every run reads the same rows in the same order. Because every microbatch must have room for the longest row, ATTUNE measures the dataset's token-length distribution before tuning, and the agent reads the resulting layout limits before its first hypothesis.

\paragraph{Scoring.} Each run lasts a fixed number of optimizer steps, set by the operator. The first steps absorb startup and compilation and are discarded, and the rest form the measured window $W$. We score a recipe $x$ on the set $G$ of its GPUs by its aggregate throughput over the window, where $\text{tokens}_i$ counts the tokens all GPUs process in step $i$:
\[
s(x) = \frac{\sum_{i \in W}\text{tokens}_i}{|G|\,\sum_{i \in W}\text{stepTime}_i}
\]
and the search maximizes $s(x)$. Summing before dividing keeps short steps from weighing more than long ones.

\paragraph{Baseline and band.} Throughput varies from run to run, so with one run per candidate, the best of many equally fast recipes would still look a few percent faster than the rest. A gain therefore counts only when it exceeds a noise band, 3\% by default, measured against a run of the baseline. A candidate replaces the leader only when it beats it by more than the band; a result inside the band is a tie and is not rerun.

\paragraph{Control plane.} The control plane is a set of deterministic scripts that launch the agent's candidates and record their results. To launch a candidate, a run generator copies the baseline job and changes only the flags and allowlisted environment variables under test, so the rest of the job stays identical. An accounting script charges every job, including failed runs and correctness checks, against the budget, and a new launch starts only if its expected cost fits what remains. The control plane also refuses to launch a recipe that has already run, so after a failure the agent must diagnose the cause and change the recipe rather than retry it.

\subsection{Correctness Gate}\label{sec:2.3}

The gate uses two checks to test that a tuned recipe trains the same objective as its baseline. Before launch, a \emph{configuration check} compares the candidate's resolved settings with the baseline's. After the run, a \emph{loss check} compares their losses at the first training step. The gate is designed never to reject a recipe that keeps the objective, since a false reject throws away a correct speedup. As a result, a fault too subtle to show in a short run can pass.

\paragraph{Configuration check.} The control plane compares the candidate's resolved arguments and environment with the baseline's. It uses resolved values, not the flags as written, because one flag can silently change others. It looks up each change in a registry that neither the agent nor any other tuner can edit and sorts it into one of three classes: known invalid, known valid and unknown. A \emph{known invalid} change alters what is trained: the data, the batch, the optimizer or its schedule, a loss term, routing or the precision format. Changes that trigger a known defect in the model code or the training image also count as known invalid (Table~\ref{tab:B1}). A \emph{known valid} change alters only how the same computation runs: a placement change, a packing change (how rows fill microbatches) or a same-math swap that passes its test. A \emph{same-math swap} computes the same function by another path, such as a fused kernel. Its \emph{component test} compares it with an FP32 reference on identical inputs; a failing swap counts as a defect. An \emph{unknown} change is any other change. The configuration check refuses known invalid changes, passes known valid ones, and lets unknown ones through to the loss check, listing them in the gate report for review.

\paragraph{Loss check.} The candidate's \emph{step-0 loss}, its loss at the first training step, must lie within 5\% of the baseline's, and all logged losses must be finite. Step 0 applies the same loaded weights to the same first batch before any update. A legitimate change can therefore move this loss only through floating-point evaluation order, by less than 2\% in our runs. Faults that load the wrong weights or change the forward pass, such as labels offset by one token, move it by more than 150\% (Section 3.5).

The control plane ships the fastest candidate that passes the gate and beats the baseline by more than the band; if none does, it ships the baseline. We also run Megatron-Bridge's convergence check \citep{nvidia2026megatronbridge} on the loss curves, but only as a report: on GLM-4.7-Flash it rejects 151 of 702 pairs of runs of the same recipe (Appendix B).

\subsection{Harness}\label{sec:2.4}

\paragraph{Autonomy.} ATTUNE builds on the \emph{persistent goal} feature of coding-agent harnesses such as Claude Code \citep{anthropic2026claude}, which keeps one task active across many agent turns. A stop hook blocks the agent from ending its turn until the campaign has spent its budget or recorded a stop with evidence. When the agent tries to ask the user a question, a pre-tool hook tells it to take the default and record the assumption. A launcher starts the campaign with one command and resumes the session after a crash.

\paragraph{Durable state.} An append-only ledger is the campaign's source of truth. Each row records one run, from its recipe and diagnosis to its verdict and GPU-hours. A fresh session rebuilds the campaign from the ledger and continues.

\paragraph{Self-evolution.} During a campaign, when a run establishes a generalizable fact about a lever, the agent writes it down as a lesson, with the evidence and the workload behind it. At the end of the campaign, the agent also writes a short summary for later campaigns to read. Lessons enter the lever catalog only after manual review (Section~\ref{sec:6.1}).

\section{Evaluation}\label{sec:3}

We evaluate ATTUNE on two sets of public recipes and in Amazon's production post-training jobs. We first measure its gains over the public recipes (Section~\ref{sec:3.2}) and compare it with OptiCo, Bayesian optimization and random search (Section~\ref{sec:3.3}). We then run extensive ablation studies of ATTUNE's design choices (Section~\ref{sec:3.4}), test the correctness gate (Section~\ref{sec:3.5}) and report the production results (Section~\ref{sec:3.6}).

\subsection{Setup}\label{sec:3.1}

\paragraph{Models.} The \emph{reproduced} set holds four of NVIDIA's NeMo AutoModel benchmark recipes, run unchanged in NVIDIA's container on the H100s they were written for \citep{nvidia2026nemoautomodel}: Qwen3-30B-A3B \citep{qwen2025qwen3}, GPT-OSS-20B \citep{openai2025gptoss120b}, Nemotron-3-Nano \citep{nvidia2025nemotron} and Llama-3.1-70B \citep{llama2024llama}. Three of these models are not held out: ATTUNE was developed on Qwen3-30B-A3B and GPT-OSS-20B, and its released catalog includes lessons from the Nemotron-3-Nano ablations of Section 3.4. The \emph{ported} set holds the four held-out models, run on slime \citep{zhu2025slime} and Megatron-LM: a dense 405B model (Llama-3.1-405B) and three MoE models from 141B to 355B parameters (Mixtral-8x22B, MiniMax-M2, GLM-4.5). For each model we take its \emph{strongest public recipe} (Table~\ref{tab:A3}): a vendor throughput benchmark for the hardware where one exists, and otherwise the framework's own recipe.

\paragraph{Workload.} Each public-recipe campaign keeps its recipe's sequence length, global batch, precision and GPU count, so a gain is measured against the recipe at the shape it was written for. The one exception is MiniMax-M2, whose 128-GPU recipe we port to 64 GPUs. The ported set starts from the released weights and trains on real text, while the reproduced set keeps NVIDIA's benchmark workload: mock tokens, balanced expert routing and, except on Llama-3.1-70B, randomly initialized weights. A step-0 loss check on mock tokens says little about correctness, so we also train each shipped recipe and its public recipe side by side on real text from the released weights.

\paragraph{Baseline and tuners.} Every tuner starts from the same baseline, the public recipe on that workload, which in the ported set includes the porting changes of Appendix A. ATTUNE runs a fixed copy of its skill, without the summaries of past campaigns. The other tuners keep their own algorithms; OptiCo keeps its analyzer, prompts and schema. In the ported set, all agents use Claude Opus 4.8 with web tools disabled. Each campaign runs in its own sandbox and reaches the cluster only through a broker (Appendix C). The reproduced set uses a lighter protocol, with one campaign per model and a per-model budget (Table~\ref{tab:1}). Its agent, Claude Opus 5.5 with web tools enabled, runs the released skill through an AutoModel adapter, without the sandbox or broker. We score its runs over the benchmark's own window, with no final re-measurement.

\begin{table*}[!tp]
\caption{ATTUNE speeds up the strongest public recipe of seven of the eight models, by 7--26\% on NVIDIA's benchmark recipes and by up to 2.6$\times$ on held-out models; on Mixtral-8x22B it keeps the public recipe. Each campaign keeps the recipe's workload and GPU count (MiniMax-M2's 128-GPU recipe runs on 64). NVIDIA recipes have one campaign each; held-out values are means over three campaigns. MFU: model FLOPs utilization (Appendix D). GPU-h is per campaign, with the budget in parentheses. Tables~\ref{tab:D2} and~\ref{tab:D3} give every held-out campaign.}\label{tab:1}
\vskip 0.1in
\small
\begin{tabularx}{\linewidth}{@{}>{\hsize=1.55\hsize\raggedright\arraybackslash}X>{\hsize=0.78\hsize\raggedright\arraybackslash}X>{\hsize=0.95\hsize\raggedleft\arraybackslash}X>{\hsize=0.88\hsize\raggedleft\arraybackslash}X>{\hsize=0.80\hsize\raggedleft\arraybackslash}X>{\hsize=0.95\hsize\raggedleft\arraybackslash}X>{\hsize=0.80\hsize\raggedleft\arraybackslash}X>{\hsize=1.29\hsize\raggedleft\arraybackslash}X@{}}
\toprule
\addlinespace[3pt]
 &  & \multicolumn{2}{c}{Throughput (tokens/s/GPU)} &  & \multicolumn{2}{c}{MFU (\%)} &  \\
\cmidrule(lr){3-4}\cmidrule(lr){6-7}
Model & GPUs & Public recipe & ATTUNE & Gain & Public recipe & ATTUNE & GPU-h (budget) \\
\midrule
\addlinespace[4pt]
\multicolumn{8}{l}{\emph{NVIDIA benchmarks, reproduced on NeMo AutoModel}} \\
\addlinespace[4pt]
Qwen3-30B-A3B & 8$\times$H100 & 11,963.1 & 13,151.1 & +9.9\% & 27.9 & 30.7 & 37.2 (70) \\
GPT-OSS-20B & 8$\times$H100 & 12,970.3 & 15,222.5 & +17.4\% & 30.1 & 35.3 & 18.1 (65) \\
Nemotron-3-Nano & 8$\times$H100 & 16,538.1 & 17,726.7 & +7.2\% & 33.4 & 35.8 & 23.5 (65) \\
Llama-3.1-70B & 64$\times$H100 & 801.2 & 1,012.5 & +26.4\% & 36.4 & 46.0 & 133.4 (500) \\
\midrule
\addlinespace[4pt]
\multicolumn{8}{l}{\emph{Held-out models, ported to slime and Megatron-LM}} \\
\addlinespace[4pt]
MiniMax-M2 & 64$\times$B200 & 895.2 & 2,365.1 & +164.2\% & 3.2 & 8.5 & 203.5 (530) \\
GLM-4.5 & 128$\times$B200 & 723.6 & 969.0 & +33.9\% & 7.2 & 9.7 & 669.1 (1,880) \\
Llama-3.1-405B & 128$\times$B200 & 394.9 & 421.9 & +6.8\% & 22.2 & 23.7 & 393.9 (1,850) \\
Mixtral-8x22B & 128$\times$H100 & 1,072.4 & 1,072.4 & 0.0\% & 26.3 & 26.3 & 173.4 (740) \\
\bottomrule
\end{tabularx}
\end{table*}

\paragraph{Budget, scoring and comparisons.} In the ported set every tuner gets the same GPU-hour budget per model: 30 times the cost of one baseline benchmark run (Table~\ref{tab:D1}). Each benchmark run lasts 14 optimizer steps, scored over steps 4--13. Reported throughputs come from a final interleaved re-measurement of the baseline and every shipped recipe. To measure the noise, we run each baseline at least five times on different node allocations. Their mean is the baseline throughput that search compares candidates against. The noise band is 3\% or three times their relative standard deviation, whichever is larger; only Llama-3.1-405B's band exceeds 3\% (3.88\%, Table~\ref{tab:D1}). Comparisons between tuners use the \emph{mean rule}: tuner A beats tuner B when the mean throughput of A's shipped recipes exceeds B's by more than the model's noise band. Each comparison also reports an exact permutation p-value (Appendix D).

\subsection{ATTUNE against the Strongest Public Recipes}\label{sec:3.2}

Table~\ref{tab:1} compares the throughput of each public recipe and its ATTUNE-tuned recipe, at the public recipe's own workload and GPU count. ATTUNE improves throughput in both the reproduced and the ported set.

\paragraph{Reproduced recipes.} We first reproduce NVIDIA's four AutoModel benchmark recipes on our cluster. Three come within 2\% of NVIDIA's published throughput, and Llama-3.1-70B within 8\%. Starting from these reproductions, ATTUNE raises throughput by 7--26\%. The largest gain is on Llama-3.1-70B, where ATTUNE drops context parallelism and switches attention to FlashAttention-3 \citep{shah2024flashattention}, after which compute takes 98.5\% of the step. Qwen3-30B-A3B gains from switching the kernel of the token dispatcher, which sends tokens to experts, and from changing how many streaming multiprocessors (SMs) the dispatcher reserves. Nemotron-3-Nano gains from the dispatcher's kernel alone, and GPT-OSS-20B from fused attention, the dispatcher's SM count and the expandable-segments setting of PyTorch's memory allocator. On real text from the released weights, the shipped changes pass Megatron-Bridge's convergence check against the public recipe (Appendix D).

\paragraph{Ported recipes.} On the models whose recipes we ported to our stack, ATTUNE raises throughput by 164.2\% on MiniMax-M2, 33.9\% on GLM-4.5 and 6.8\% on Llama-3.1-405B. On Mixtral-8x22B its best candidate stays inside the noise band, so it keeps the public recipe. The gain is largest where the recipe was written for another setting. MiniMax-M2's recipe targets reinforcement learning on 128 GPUs and keeps the optimizer state in CPU memory, and GLM-4.5's targets H100, leaving room to tune on B200. The Llama-3.1-405B and Mixtral-8x22B recipes are vendor benchmarks tuned for the hardware they run on, so they leave less to gain from configuration. On our stack they reach 58--60\% of the vendors' published throughput, which the vendors measure in their own training loops (Table~\ref{tab:A3}). About 30\% of our extra time per step, relative to the vendors' runs, is slime's own work between training steps; the rest is inside the steps. On MiniMax-M2, all three campaigns turn the optimizer CPU offload off and raise pipeline parallelism. On GLM-4.5, ATTUNE finds the step bound by communication and kernel launches. All three campaigns move the expert all-to-all inside a node, and the fastest also sets TP = 1.

\subsection{Tuner Comparison}\label{sec:3.3}

We compare ATTUNE with an LLM-agent tuner and three black-box searches on MiniMax-M2 and GLM-4.5. Every tuner gets the same GPU-hour budget, correctness gate and shipping rule, and runs three campaigns per model because a campaign is stochastic.

\begin{table*}[!tp]
\caption{ATTUNE ships faster recipes than every other tuner on MiniMax-M2 / 64$\times$B200 and GLM-4.5 / 128$\times$B200, with the fewest GPU-hours and launches. Values are means over three campaigns, with the standard deviation for gain over the public recipe. Bold marks each model's best value. ATTUNE's gain reaches p = 0.05, the lowest possible with three campaigns per tuner, against every tuner except random search on GLM-4.5 (p = 0.10). Appendix D defines launches, failed launches and wall-clock. BO: Bayesian optimization; TPE: tree-structured Parzen estimator.}\label{tab:2}
\vskip 0.1in
\small
\begin{tabularx}{\linewidth}{@{}>{\hsize=0.894\hsize\raggedright\arraybackslash}X>{\hsize=1.735\hsize\raggedright\arraybackslash}X>{\hsize=1.095\hsize\raggedleft\arraybackslash}X>{\hsize=0.745\hsize\raggedleft\arraybackslash}X>{\hsize=0.670\hsize\raggedleft\arraybackslash}X>{\hsize=1.117\hsize\raggedleft\arraybackslash}X>{\hsize=0.745\hsize\raggedleft\arraybackslash}X@{}}
\toprule
Model & Tuner & Gain (\%) & GPU-h used & Launches & Failed launches & Wall-clock \\
\midrule
MiniMax-M2 & ATTUNE & \textbf{+164.2} $\pm$ 12.5 & \textbf{203} & \textbf{13.3} & 23\% & 3.1 h \\
 & OptiCo (LLM agents) & +22.4 $\pm$ 1.9 & 386 & 26.0 & 24\% & 6.9 h \\
 & BO (TPE) & +46.6 $\pm$ 36.6 & 492 & 35.0 & 53\% & \textbf{2.7 h} \\
 & BO + memory pre-filter & +55.5 $\pm$ 5.2 & 489 & 22.7 & \textbf{4\%} & 2.9 h \\
 & Random search & +32.7 $\pm$ 40.0 & 487 & 38.0 & 67\% & \textbf{2.7 h} \\
\midrule
GLM-4.5 & ATTUNE & \textbf{+33.9} $\pm$ 18.0 & \textbf{669} & \textbf{8.3} & 24\% & \textbf{6.4 h} \\
 & OptiCo (LLM agents) & 0.0 $\pm$ 0.0 & 1,407 & 22.7 & 19\% & 15.7 h \\
 & BO (TPE) & +1.8 $\pm$ 3.1 & 1,675 & 19.7 & 39\% & 9.6 h \\
 & BO + memory pre-filter & +3.9 $\pm$ 3.4 & 1,695 & 13.7 & \textbf{5\%} & 9.5 h \\
 & Random search & +7.8 $\pm$ 13.5 & 1,688 & 22.7 & 51\% & 10.2 h \\
\bottomrule
\end{tabularx}
\end{table*}

\paragraph{OptiCo.} We run the open-source OptiCo \citep{chen2026optico} with the same agent model as ATTUNE, through an adapter for our stack and our loop driver (Appendix E). We keep its published search space, a fixed schema of 11 knobs. The schema leaves out expert-tensor and context parallelism, virtual stages, the dispatcher and optimizer offload, so these stay at the baseline's values.

\paragraph{Black-box search.} We run three black-box searches over a fixed space of settings that keep the training objective: uniform random search, Bayesian optimization \citep[{TPE;}][]{bergstra2011algorithms,akiba2019optuna}, and Bayesian optimization with a memory pre-filter that skips configurations predicted to run out of GPU memory.

\paragraph{ATTUNE ships faster recipes than the other tuners.} On MiniMax-M2 its mean gain is about three times that of the best alternative, Bayesian optimization with the pre-filter (Table~\ref{tab:2}). On GLM-4.5 it is more than four times that of the best alternative there, random search. Every ATTUNE campaign on MiniMax-M2 beats every campaign of the other tuners. On GLM-4.5, each of its three campaigns gains at least 22\%, while eight of the other tuners' 12 campaigns ship the public recipe.

\begin{figure*}[!t]
\centering
\includegraphics[width=\textwidth]{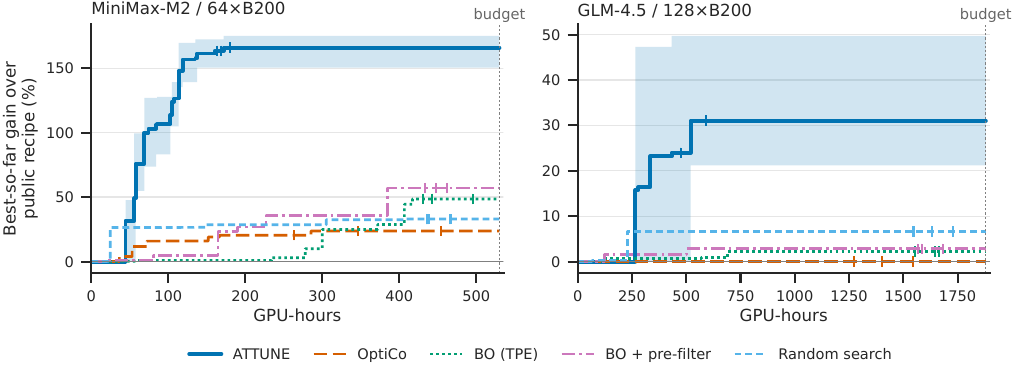}
\caption{ATTUNE finds its best recipe within about a third of the budget and stops soon after, while the other tuners search for most of the budget and end lower. Best-so-far search-time gain against GPU-hours on MiniMax-M2 (left) and GLM-4.5 (right): mean over each tuner's three campaigns (line) and the range of ATTUNE's campaigns (shading). Ticks mark each campaign's last search run, after which its curve runs flat; the dotted line is each model's budget (530 and 1,880 GPU-hours). BO: Bayesian optimization; TPE: tree-structured Parzen estimator; the pre-filter skips draws predicted to run out of memory.}\label{fig:2}
\end{figure*}

\paragraph{ATTUNE uses the least search compute.} ATTUNE needs the fewest GPU-hours and launches on both models, and the least wall-clock time on GLM-4.5 (Table~\ref{tab:2}). Within the first 30\% of the budget, every ATTUNE campaign finds a recipe more than twice as fast as the public recipe on MiniMax-M2 and more than 20\% faster on GLM-4.5 (Figure~\ref{fig:2}). The campaigns then stop well before the budget runs out, so on average ATTUNE uses about half the GPU-hours of any other tuner. On both models the GPU-hours saved by training on ATTUNE's recipe repay its tuning cost within a day, at least four times sooner than for any other tuner (Appendix D).

\paragraph{The other tuners miss winning layouts within their reach.} On MiniMax-M2, eight of the nine black-box campaigns drop the optimizer CPU offload, a change worth about a third of ATTUNE's mean gain (Table~\ref{tab:D3}). Every ATTUNE campaign combines that change with a deeper pipeline (PP = 4 or 8) and with TP = 1, more rows per microbatch, or both. On GLM-4.5, ATTUNE's fastest layout without its virtual stages lies inside OptiCo's schema and runs within 2\% of ATTUNE's recipe, yet OptiCo never reached it. OptiCo tried TP = 1 only at shallower pipelines, with expert parallelism still crossing nodes, and none of its runs beat the public recipe. The black-box searches have all of GLM-4.5's winning levers in their space. Most of their draws ran out of memory or were stopped early as slower than the leader; with the pre-filter, most of the draws it let through were still stopped early. ATTUNE instead finds the bottleneck, here the cross-node expert all-to-all and the tensor-parallel communication, and changes the parallel degrees that depend on each other together, reaching a winner in eight launches or fewer; the other tuners average 14 to 23 launches per campaign there (Table~\ref{tab:2}).

In campaign 2, for example, the first candidate moved the expert all-to-all inside a node and tied the baseline. The agent concluded that the step was bound by communication and kernel launches, and that expert placement did not matter. TP = 1 then ran out of memory at both 8 and 16 pipeline stages. The agent traced this to unsharded sequence activations and added full recomputation, expecting it to cost little because the GPUs sat idle between kernel launches. This layout became the leader. A later candidate that returned to PP 8 and EP 16, which puts the all-to-all back across nodes, ran about 38\% slower before the agent stopped it early. The agent concluded that the faster step exposed the cross-node all-to-all that the slow baseline step had hidden, so expert placement now mattered (Appendix F).

\begin{figure*}[!tp]
\centering
\includegraphics[width=\textwidth]{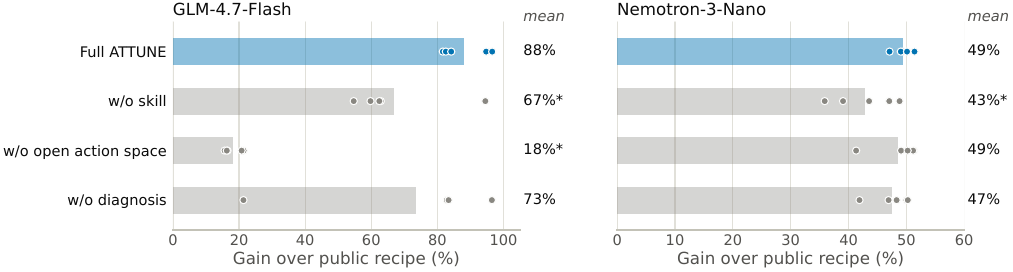}
\caption{Removing the skill lowers ATTUNE's gain on both models, and restricting the agent to OptiCo's 11 knobs lowers it on GLM-4.7-Flash, whose largest win lies outside them. Without diagnosis (no logs or profiles), the mean gain is lower on both models, but not significantly. GLM-4.7-Flash and Nemotron-3-Nano on 8$\times$H200, five campaigns per variant. Bars show each variant's mean gain and dots each campaign's gain over its own run of the public recipe; the two panels use different scales. * marks p < 0.05 in a one-sided exact permutation test of the mean gain against full ATTUNE. Table~\ref{tab:ablations} gives each p-value, failed launches and GPU-hours to a 20\% gain.}\label{fig:ablations}
\end{figure*}

\subsection{Ablations}\label{sec:3.4}

To better understand what each of ATTUNE's design choices contributes, we remove or replace one at a time and compare the result with full ATTUNE (Figure~\ref{fig:ablations}). The \emph{w/o skill} variant is the same agent and harness without the skill. The \emph{w/o open action space} variant may change only the 11 knobs of OptiCo's schema. The \emph{w/o diagnosis} variant sees each run's score and status but not its logs or profiles. The ablations use GLM-4.7-Flash \citep{zhipu2026glm47flash} and Nemotron-3-Nano \citep{nvidia2025nemotron}, both new to ATTUNE when the ablations ran. Each is trained with SFT on the miles framework \citep{radixark2026miles} on 8$\times$H200, starting from the recipe miles publishes for it. Every variant runs five campaigns per model under the same GPU-hour budget, and each campaign sees only its own records. Full ATTUNE has the highest mean gain on both models.

\paragraph{The skill.} Without the skill, the agent's mean gain falls from +88.0\% to +66.9\% on GLM-4.7-Flash and from +49.4\% to +42.9\% on Nemotron-3-Nano (p = 0.016 and 0.008). It also fails far more of its launches and needs up to 2.3 times the GPU-hours to reach a 20\% gain (Table~\ref{tab:ablations}).

\paragraph{The open action space.} Restricted to OptiCo's knobs, the agent falls far behind on GLM-4.7-Flash (p = 0.004) and ties on Nemotron-3-Nano. The largest win on GLM-4.7-Flash, turning optimizer CPU offload off, is not in the schema, while on Nemotron-3-Nano the largest wins lie inside it.

\paragraph{Diagnosis.} Without logs and profiles, the agent's mean gain is lower than full ATTUNE's on both models: +73.4\% against +88.0\% on GLM-4.7-Flash and +47.5\% against +49.4\% on Nemotron-3-Nano. With five campaigns per variant, neither difference is significant (p = 0.24 and 0.14).

\subsection{The Correctness Gate}\label{sec:3.5}

To check that the gate passes legitimate changes and catches faults, we test it on two sets of runs. The first set holds runs whose correct verdict we know in advance (Table~\ref{tab:4}): repeats of public recipes, legitimate recipes whose step-0 numerics differ from the baseline's, injected faults, and the ablation runs of Section 3.4. The second holds every candidate of the evaluation and ablation campaigns.

\paragraph{Known-answer runs.} Table~\ref{tab:4} shows that the gate passes all 460 legitimate runs and that its two checks catch different faults. The configuration check refuses all 29 configuration faults. Faults that change the forward pass, such as labels offset by one token, raise the step-0 loss by more than 150\%, so the loss check rejects all five. The gate misses all nine remaining faults. Faults that act only through the optimizer update, such as a skipped gradient reduction, leave the step-0 loss unchanged. A loss mask shifted by one token does change it, but only by 2.0\%, about as much as the largest legitimate change. These code faults matter less for ATTUNE, which changes configuration, not code. When a setting can cause the same fault, such as a changed auxiliary-loss coefficient, the configuration check refuses that setting. Two real stack defects show why the configuration check needs a known-defect list: EP = 1 with an expert-tensor parallelism (ETP) of 4 or more, and PP = 32. Both move the step-0 loss by 0.02\% or less, so the loss check misses them.

\begin{table}[!h]
\caption{The correctness gate passes every legitimate run and catches every configuration fault and forward-pass code fault. It misses update-path faults, which leave the step-0 loss within tolerance. A legitimate run repeats a recipe or changes only settings the configuration check lists as keeping its math, such as placement or packing; 132 of them also pass Megatron-Bridge's convergence check over 20 steps. Tables~\ref{tab:B1} and~\ref{tab:B2} list the faults.}\label{tab:4}
\vskip 0.1in
\small
\setlength{\tabcolsep}{3pt}
\begin{tabular*}{\linewidth}{@{\extracolsep{\fill}}lrl@{}}
\toprule
Known-answer runs & Correct verdicts & Decided by \\
\midrule
Legitimate runs & 460 of 460 & both checks pass \\
Configuration faults & 29 of 29 & configuration check \\
Forward-pass code faults & 5 of 5 & loss check \\
Update-path code faults & 0 of 9 & neither (missed) \\
\bottomrule
\end{tabular*}
\end{table}

\paragraph{In the campaigns.} In the ported-set campaigns of Sections 3.2--3.3, every candidate that reached the gate passed. Their step-0 shifts were at most 0.03\%, far inside the 5\% tolerance, so the loss check never came close to rejecting a legitimate candidate. In the ablation campaigns of Section 3.4, the gate refused 20 candidates with known defects: 19 uneven pipeline splits on Nemotron-3-Nano and one fused router on GLM-4.7-Flash. The splits ran before the registry listed them; reported results apply the gate of Section~\ref{sec:2.3}. In one campaign the fastest run was such a split, so that campaign reports the next fastest candidate. To check that a recipe that passes the gate also trains correctly over a longer run, we trained one shipped recipe for 1,000 steps. Its loss stayed within run-to-run noise of the public recipe's (Appendix B).

\subsection{Deployment}\label{sec:3.6}

ATTUNE is deployed in Amazon's production post-training, where its recipes train models that power Amazon's Ads Agent \citep{amazon2026ads}, which turns hours of advertising-campaign setup and targeting into minutes. Ads Agent is one of the AI-powered tools of Amazon Ads, a business that reported \$19.8 billion in revenue in Q2 2026. Figure~\ref{fig:gains} reports three of the models ATTUNE tunes, which are also the models it was developed on (Appendix~\ref{sec:G}). Each ATTUNE recipe replaced the job's previous recipe, tuned by our team's experts, and matched its loss over 100 training steps. Measured on 8,192-token rows at global batch 128, the tuned recipes raise throughput by a geometric mean of 1.7$\times$ and up to 3.9$\times$. The largest gains come from changing how a recipe trades memory for speed. On B200, for example, the Qwen3.5-397B recipe drops CPU offload, adds virtual stages and runs on half the GPUs. Its throughput per GPU rises 3.9$\times$, and the job's total throughput about 2$\times$.

\section{An Open Leaderboard of Tuned Recipes}\label{sec:4}

Few open training recipes are tuned, and fewer report their throughput. The published figures cover a handful of model and accelerator pairs, each with its own workload and metric, so a practitioner cannot tell which recipe is fastest. We release OpenRecipes (\href{https://openrecipes.ai}{\texttt{openrecipes.ai}}), an open leaderboard of tuned training recipes. A recipe is compared only with recipes measured under the same protocol, which fixes the workload and the scored steps. The current protocol uses 8,192-token SFT rows at global batch 128 in BF16, scored over steps 5--19 of a 20-step run. Each entry lists its tokens/s/GPU, its cost per million tokens, and the exact recipe and framework versions.

An entry may change only what the configuration check marks as known valid: placement, packing and tested same-math swaps. It must also pass Megatron-Bridge's convergence check against the recipe it was tuned from. OpenRecipes now lists more than 60 recipes for more than 20 models. Its entries are seeded from ATTUNE campaigns, and the leaderboard is open to submissions. A team submits a recipe through a pull request, and we review its measurement and correctness evidence before listing it. Teams can adopt a listed recipe instead of repeating the search and use its cost to weigh in-house training against a paid service.

\section{Related Work}\label{sec:5}

\paragraph{Agentic LLM tuners.} The systems closest to ATTUNE are OptiCo \citep{chen2026optico}, PROMPTS \citep{ding2026prompts} and ASAP \citep{ding2025asap}. OptiCo's LLM agents launch training jobs, read the results and propose the next configuration. PROMPTS and ASAP read the profile of an existing run and propose sharding configurations for an engineer to run. CCL-Search \citep{ding2026cclbench} runs a similar loop with a single LLM over knobs the user lists. Megatron-Bridge's agent skills \citep{nvidia2026megatronbridge} guide profiling-driven tuning in an interactive session. Concurrent with our work, Primus \citep{raina2026primus} uses two LLMs to search Megatron-LM's parallel layout, recomputation, precision and MoE backend, and scores candidates with a projection model instead of runs at the target scale.

ATTUNE shares OptiCo's run-and-measure loop and differs from these systems in four ways. First, it is a single agent under a skill, while OptiCo coordinates four agents and PROMPTS and ASAP use a three-agent pipeline. Prior work finds that a single agent can match multi-agent discussion \citep{wang2024rethinking}, while multi-agent systems add coordination failures \citep{cemri2025multiagent} and cost \citep{smit2024going}. Second, it can change any framework argument that its gate does not mark as changing the objective, while PROMPTS and ASAP tune only the parallelism mapping and OptiCo an 11-knob schema. Third, it checks that a change preserves the training objective: OptiCo and CCL-Search check that a configuration runs, PROMPTS that it compiles, Primus that it fits in memory, and the Megatron-Bridge skills ask the agent itself to compare losses. Fourth, it ships a recipe with no human step in the search and carries lessons across campaigns. PROMPTS returns three recommendations for a human to review, and OptiCo's memory lasts one optimization. We run only OptiCo as a baseline, since PROMPTS and ASAP release no code and target TPUs.

\paragraph{Planners and autoconfigurators.} Planners search the parallelization space with a cost model or simulator. SimAI \citep{wang2025simai} and FlexFlow \citep{jia2019beyond} estimate a plan's step time without running it. Alpa \citep{zheng2022alpa}, Galvatron \citep{miao2022galvatron} and Metis \citep{um2024metis} synthesize plans under memory constraints. Sailor \citep{strati2025sailor} and AutoScout \citep{shong2026autoscout} extend planning to heterogeneous clusters and to runtime parameters, and Aceso \citep{liu2024aceso} iteratively relieves the predicted bottleneck. DeepSpeed's autotuner \citep{deepspeed2021deepspeed} and NeMo's AutoConfigurator \citep{nvidia2026nemo} sweep a bounded set of documented knobs. AutoConfigurator searches a grid of only parallel degrees and microbatch size. Both tools search a space fixed in advance, mostly placement, while ATTUNE can change any framework argument. On MiniMax-M2 and GLM-4.5 its recipes also swap the token dispatcher and change optimizer offload \citep{ren2021zerooffload} and communication overlap, which lie outside a placement planner's space.

\paragraph{Checking tuned configurations.} TrainVerify \citep{lu2025trainverify} formally verifies that a parallel plan is equivalent to the model's logical specification. TTrace \citep{jiang2025ttrace} compares a distributed run's tensors with a reference implementation, and CATWILD \citep{cano2026catwild} compares the outputs of tuned compiler configurations with those of the defaults. ATTUNE's gate needs only the starting recipe's run as its reference.

\section{Discussion}\label{sec:6}

\subsection{Lessons Learned}\label{sec:6.1}

\paragraph{The catalog needs curation before reuse.} Our past campaigns wrote their lessons into the catalog as general rules. When we later checked the catalog against the campaigns' own records, about a fifth of its lessons were wrong, and several held only on the workload they were measured on (Table~\ref{tab:G1}). On GLM-4.5, one campaign followed a wrong entry about turning tensor parallelism off and never reached the winning layout. Each campaign therefore records its lessons separately from the catalog, with the runs that support each lesson and the workload it was measured on, and lessons enter the catalog only after review. Many correct entries were also redundant. Asked without the skill, the current model already knew 19 of the 106 curated entries fully and 59 in part. For about a quarter of the curated entries, however, the model's answer contradicted our measurements. For example, it expects fusion and overlap flags to add up to a large gain, while in our campaigns they tied or lost. A catalog helps only where it corrects the model or states facts about the stack that the model cannot know, so we keep only the entries the current model misses.

\paragraph{The skill's search rules need enforcement.} When a layout runs out of memory, the skill tells the agent to retry it with its memory-saving settings turned back on before ruling it out. On GLM-4.5, only campaign 2 retried TP = 1 at 16 pipeline stages with full recomputation, and it shipped the best recipe (+54.7\%). The other two did not and shipped gains of +22.4\% and +24.7\% (Table~\ref{tab:D2}). The harness already enforces the correctness gate and the budget outside the agent, so the next step is to move the search rules whose violations cost the most into the harness as well.

\subsection{Limitations}\label{sec:6.2}

\paragraph{Priors.} We distilled the lever catalog from past campaigns. One held-out model, GLM-4.5, belongs to the same family as the development model GLM-5.2. Knowledge carried over from past campaigns could account for part of ATTUNE's gains on GLM-4.5 and on the three reproduced-set models it was developed or ablated on (Section~\ref{sec:3.1}).

\paragraph{Gate coverage.} The gate cannot catch every fault: it misses faults that act only through the optimizer update (Section 3.5) and unknown flags that break training without changing the step-0 loss. Later-step losses and the step-0 gradient norm do not catch them either (Appendix~\ref{sec:B}). The gate report names every unknown flag in a shipped recipe so a person can review it. A recipe should still pass a full training run before production.

\paragraph{Scope.} ATTUNE tunes the configuration that the framework and the training image expose and leaves framework code, kernels and the image unchanged. It covers only the training step, so the rollout and generation phases of reinforcement learning lie outside its scope. The recipe it returns is the best its search found for that cluster and stack, with no guarantee of optimality or of transfer to another setting.

\section{Conclusion}\label{sec:7}

ATTUNE is an agentic tuner that speeds up multi-node LLM training recipes unattended. A single coding agent, guided by a tuning skill, runs the search, and a correctness gate outside the agent checks each recipe. ATTUNE's recipes raise the per-GPU throughput of the Amazon production jobs we report by 1.7$\times$ on average, speed up NVIDIA's benchmark recipes and held-out public recipes, and beat those of Bayesian optimization, random search and OptiCo. We release ATTUNE and OpenRecipes, an open leaderboard of more than 60 tuned and verified recipes, so that teams can start from a fast recipe that trains the same model instead of repeating the search.\attunemark{mainend}

\bibliography{refs}
\bibliographystyle{mlsys2025}

\section*{AI Usage Disclosure}

ATTUNE is itself an AI system, so generative AI is part of the work under study. The ported-set campaigns of ATTUNE and OptiCo and the ablation variants ran Claude Opus 4.8 through Claude Code, as described in Sections 2--3. The campaigns of the reproduced set (Section 3.1) ran Claude Opus 5.5 through Claude Code. The catalog check in Section 6.1 queried Claude Opus 5.5. Beyond the system under study, we used Claude Code with Claude Opus models as follows.

\paragraph{Ideation and experiment design.} The authors conceived ATTUNE, designed its architecture and correctness gate, and set the evaluation protocol, budgets and decision rules. AI-simulated reviewer reads were used to find weaknesses in the design and evaluation; the authors decided which to address.

\paragraph{Code.} Much of ATTUNE's harness, skill scripts, correctness tools, baseline adapters, and analysis and plotting code was written with Claude Code from the authors' specifications. The authors reviewed and tested every change, and an independent agent also reviewed changes to measurement and scoring.

\paragraph{Experiments and analysis.} Agents ran post-hoc analyses under the authors' direction: the re-measurement sensitivity check, the transcript and ledger analyses, and the catalog audit. The authors checked their results against the raw records.

\paragraph{Data.} We generated no training data with a model. Real-text rows come from public corpora and are processed by scripts. The reproduced set's benchmark rows are the benchmark's own mock tokens.

\paragraph{Literature search.} Agents with web access assisted the literature search. Every reference was checked against its canonical record (arXiv, Crossref or the publisher's page), first by scripted and agent-based checks and then by hand by the authors. Agents also checked each claim the paper attributes to a cited work against that work.

\paragraph{Writing.} The authors set the structure and content of every section, wrote parts of the text themselves, and revised the whole paper over many rounds. Claude Code drafted much of the text from the authors' outlines and applied their revisions. The authors decided every claim and every edit, and take responsibility for the content.

\ifdefined\attunearxiv\else\clearpage\fi%
\appendix
\renewcommand{\thetable}{\thesection\arabic{table}}
\renewcommand{\thefigure}{\thesection\arabic{figure}}
\renewcommand{\theHtable}{\thesection.\arabic{table}}

\section{Workloads, Baselines and Datasets}\label{sec:A}
\setcounter{table}{0}

Each held-out model trains on real text, packed to its recipe's sequence length: FineWeb-Edu \citep{penedo2024fineweb} for MiniMax-M2 and GLM-4.5, and PG-19 books \citep{rae2020compressive} for Llama-3.1-405B and Mixtral-8x22B, each with the model's own tokenizer. A fidelity check runs each model in our stack and in its released implementation on the same rows. Their losses agree within 1\%, and for the MoE models the per-layer expert choices agree on at least 95\% of assignments.

\paragraph{OpenRecipes workload.} The current OpenRecipes protocol (Section~\ref{sec:4}) uses 8,192-token rows of SFT conversations from the Tulu 3 SFT mixture \citep{lambert2025tulu}, rendered with each model's chat template, at global batch 128 in BF16.

\paragraph{Stack fixes.} Six stack fixes apply to all tuners:
\begin{enumerate}
\item A fix for slime's re-entry bug under data-parallel overlap, with Megatron's FP8 parameter-sync backport.
\item The userbuffers bootstrap for tensor-parallel communication overlap.
\item Skipping slime's per-step copy of the weights to host memory, which only reinforcement-learning rollouts need.
\item Fixed pipeline-communication shapes when microbatches are single fixed-length rows.
\item Garbage collection every 50 steps instead of every step.
\item Restoring NCCL's default 4 MiB buffer size.
\end{enumerate}
Each fix passes a test showing that it leaves the loss unchanged within run-to-run noise.

\paragraph{Changes from the public recipes.} Table~\ref{tab:A2} lists each model's changes. The models run under slime's training loop with a pretokenized data feed and the stack fixes above.

\begin{table*}[!tp]
\caption{Changes from each public recipe that affect throughput or the baseline. The baseline every tuner starts from includes them. All models start from released weights.}\label{tab:A2}
\vskip 0.1in
\small
\begin{tabularx}{\linewidth}{@{}llX@{}}
\toprule
Model & GPUs & Change from the public recipe \\
\midrule
MiniMax-M2 & 64$\times$B200 & SFT on fixed 8,192-token FineWeb-Edu rows instead of the recipe's reinforcement learning \\
 & & 64 GPUs instead of 128, keeping tensor, pipeline, expert-tensor and context parallelism \\
 & & Released FP8 weights dequantized to BF16 \\
 & & EP 4$\rightarrow$8 to fit in memory \\
 & & QK-norm weight gradients averaged, not summed, over tensor-parallel ranks \\
\addlinespace[4pt]
GLM-4.5 & 128$\times$B200 & Multi-token-prediction layer removed \\
 & & B200 instead of the recipe's H100, with the published layout unchanged \\
 & & Data-parallel overlap kept on as published, through a stack fix \\
\addlinespace[4pt]
Llama-3.1-405B & 128$\times$B200 & Per-token loss \\
 & & Vocabulary unpadded to match the converted checkpoint \\
 & & Data- and tensor-parallel overlap kept on as published, through stack fixes and a gloo bootstrap \\
 & & Learning rate scaled by 1/30 for the released weights \\
\addlinespace[4pt]
Mixtral-8x22B & 128$\times$H100 & RoPE base 10\textsuperscript{6} as in the released model, which the recipe omits \\
 & & Selective recomputation of the expert activation to fit in memory; the recipe recomputes nothing \\
 & & Userbuffers bootstrap over gloo \\
\bottomrule
\end{tabularx}
\end{table*}

\begin{table*}[!tp]
\caption{Public recipes and their workloads, with 8 GPUs per node (B200: 180 GB and H100: 80 GB of HBM per GPU). Throughput is in tokens/s/GPU, except for Mixtral-8x22B ($\dagger$: TFLOP/s/GPU). ``Published'' is the throughput the recipe's authors report \citep{nvidia2026megatronbridge,yan2026megatron,nvidia2026nemoautomodel}; against these figures, ATTUNE's recipes for the four NVIDIA benchmarks are 6--17\% faster (Table~\ref{tab:1} gives their throughput). The NVIDIA benchmarks train on mock tokens with balanced routing and random initialization, except Llama-3.1-70B, which loads the released weights. Recipe: the source of the public recipe; the MiniMax-M2 recipe is written for 128 GPUs and the GLM-4.5 recipe for H100. For the reproduced set, ``Our stack'' is the unchanged recipe in NVIDIA's container (26.08, AutoModel 922e83e4) on our cluster; for the ported set it is the public recipe on slime and Megatron-LM with the changes of Appendix A. Appendix D defines MFU; all MFU values use its FLOP count, not the vendors'. Params: total / active. Seq: sequence length. GBS: global batch size in rows. FP8-CS: FP8 with current scaling. --: none published.}\label{tab:A3}
\vskip 0.1in
\small
\setlength{\tabcolsep}{3pt}
\begin{tabular*}{\linewidth}{@{\extracolsep{\fill}}lllllrrlrrr@{}}
\toprule
Model & Type & Params & GPUs & Recipe & Seq & GBS & Precision & Published & MFU & Our stack \\
\midrule
\multicolumn{11}{@{}l}{\emph{NVIDIA benchmarks, reproduced on NeMo AutoModel}} \\
\addlinespace[2pt]
Qwen3-30B-A3B & MoE & 30B/3B & 8$\times$H100 & AutoModel & 4K & 512 & BF16 & 12,040 & 28\% & 11,963.1 \\
GPT-OSS-20B & MoE & 21B/3.6B & 8$\times$H100 & AutoModel & 4K & 256 & BF16 & 13,058 & 30\% & 12,970.3 \\
Nemotron-3-Nano & Mamba-MoE & 31.6B/3.2B & 8$\times$H100 & AutoModel & 4K & 512 & BF16 & 16,789 & 34\% & 16,538.1 \\
Llama-3.1-70B & dense & 70B & 64$\times$H100 & AutoModel pretrain & 8K & 128 & BF16 & 866.8 & 39\% & 801.2 \\
\midrule
\multicolumn{11}{@{}l}{\emph{Held-out models, ported to slime and Megatron-LM}} \\
\addlinespace[2pt]
MiniMax-M2 & MoE & 230B/10B & 64$\times$B200 & slime & 8K & 256 & BF16 & -- & -- & 895.2 \\
GLM-4.5 & MoE & 355B/32B & 128$\times$B200 & Megatron-Bridge & 4K & 2,048 & BF16 & -- & -- & 723.6 \\
Llama-3.1-405B & dense & 405B & 128$\times$B200 & Megatron-Bridge & 8K & 64 & FP8-CS & 661 & 37\% & 394.9 \\
Mixtral-8x22B & MoE & 141B/39B & 128$\times$H100 & Megatron-Core zoo & 4K & 256 & BF16 & 445$^\dagger$ & 45\% & 259.7$^\dagger$ \\
\bottomrule
\end{tabular*}
\end{table*}

\section{Correctness Gate}\label{sec:B}
\setcounter{table}{0}\setcounter{figure}{0}

This section gives the parameters of the gate's two checks (Section~\ref{sec:2.3}), the known stack defects the configuration check refuses, the gate's validation per fault, and a 1,000-step check of one shipped recipe.

\paragraph{Configuration check.} The known-invalid class of Section~\ref{sec:2.3} lists 182 arguments. Each model's registry also freezes some settings. Mixtral-8x22B and GLM-4.5 keep one row per microbatch, because more rows would change their load-balancing loss. Llama-3.1-405B keeps the published FP8 recipe, and MiniMax-M2 keeps its auxiliary-loss-free routing and constant learning rate. A \emph{component test} runs the default code path, the candidate and an FP32 reference on identical inputs at the model's shapes. It passes if routing is identical and the candidate's error against the reference is at most about twice the default path's.

\begin{table*}[!tp]
\caption{Known stack defects, refused by the configuration check for every tuner on the marked models. The first two are same-math swaps that change results on these models: fused routing changes the routing, and turning gradient-accumulation fusion off adds a BF16 rounding where the main gradients are FP32. With EP = 1 and ETP $\geq$ 4, the step-0 loss matches the baseline's and the loss departs after the first update. With more than 32 rows per microbatch, the gathered query exceeds 2\textsuperscript{31} elements and Transformer Engine's RMSNorm overflows; an uneven split loads another layer's weights into the Mamba layers. On Mixtral-8x22B, ETP = 8 with EP = 1 changes training, as it does on GLM-4.5. ETP = 4 with EP = 1 is refused on Mixtral-8x22B because of the GLM-4.5 result. EP, ETP, CP, TP, PP: expert, expert-tensor, context, tensor and pipeline parallelism.}\label{tab:B1}
\vskip 0.1in
\small
\setlength{\tabcolsep}{4pt}
\newcommand{\rh}[2]{\multicolumn{1}{c}{\footnotesize\begin{tabular}[b]{@{}c@{}}#1\\#2\end{tabular}}}
\newcommand{\y}{$\checkmark$}
\begin{tabularx}{\linewidth}{@{}l>{\raggedright\arraybackslash}Xccccc@{}}
\toprule
Known stack defect & How it shows & \rh{Mixtral}{8x22B} & \rh{}{GLM-4.5} & \rh{MiniMax}{M2} & \rh{GLM-4.7}{Flash} & \rh{Nemotron-3}{Nano} \\
\midrule
Fused routing & changes the routing & \y & \y & \y & \y & \\
Gradient-accumulation fusion off & adds BF16 rounding to FP32 gradients & \y & \y & \y & & \\
EP = 1 with ETP $\geq$ 4 & loss departs after the first update & \y & \y & \y & & \\
CP = 8 with virtual stages & hangs at step 0 & \y & & & & \\
CP $>$ 4 & a step takes about 1,800 s & & \y & & & \\
TP = 16 & cannot run & & & \y & & \\
More than 32 rows per microbatch & cannot run & & & \y & & \\
PP = 32 & halves only the load-balancing loss & & \y & & & \\
Uneven pipeline split & step-0 loss +215\% to +768\% & & & & & \y \\
\bottomrule
\end{tabularx}
\end{table*}

\begin{figure}[!htb]
\centering
\includegraphics[width=\columnwidth]{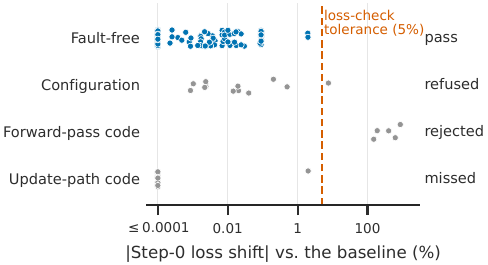}
\caption{The step-0 loss shift of every known-answer run of Table~\ref{tab:B2}, one dot per run. Shifts of 0.0001\% or less, including exact zeros, sit at the left edge. The 5\% tolerance separates the fault-free runs, all within 1.98\%, from the forward-pass code faults, all above 150\%. Configuration faults and update-path code faults mostly move the loss as little as legitimate changes do: the configuration check refuses the former before launch, and the gate misses the latter.}\label{fig:step0}
\end{figure}

\paragraph{Loss check.} The 5\% tolerance is the per-step tolerance that Megatron-Bridge's performance pipeline applies to losses between 0.5 and 2.0. Among the 328 ablation runs, the largest legitimate shift is 1.7\% on GLM-4.7-Flash and 0.02\% on Nemotron-3-Nano. Table~\ref{tab:B2} gives the other runs, and Figure~\ref{fig:step0} shows each run's step-0 shift.

\paragraph{Why only step 0.} Two cheap signals from the same short runs could, in principle, catch faults the loss check misses: the gradient norm at step 0 and the loss at later steps. We tested both on the 765 legitimate runs with saved logs (Table~\ref{tab:B2}'s fault-free runs, the ablation runs, the evaluation candidates and 25 runs on models outside this paper) and on the nine missed faults. Legitimate changes move the step-0 gradient norm by $-$68\% to +907\%, through placement, fused attention or hardware, so no threshold separates faults from legitimate runs. A 5\% rule on the step-1 loss rejects none of the 615 legitimate runs that log a comparable step 1. Beyond the faults the step-0 check already catches, it catches only EP = 1 with ETP $\geq$ 4, which the configuration check already refuses, and none of the nine missed faults: they leave the step-1 loss unchanged or within 0.05\%, except the shifted loss mask at +2.4\%, close to the largest legitimate change (2.06\%). From step 2 on, runs of the same recipe diverge, by up to 16.5\% at step 3 on GLM-4.7-Flash, so a 5\% rule would reject legitimate runs. At these steps the warmup learning rate is 1.9e-8 to 3e-5, too small for one update to move the loss.

\paragraph{Convergence check.} The gate runs Megatron-Bridge's \texttt{validate\_\allowbreak{}convergence}, copied unmodified (Table~\ref{tab:C1}), on the candidate's and the baseline's loss curves, with the thresholds of Megatron-Bridge's own performance pipeline. It lists the result in the gate report. The result is advisory only. GLM-4.7-Flash's loss spikes during its first ten steps, and 151 of the 702 ordered pairs of its 27 baseline-recipe runs fail the check against each other. A rule on the curve would therefore reject correct recipes.

\begin{table*}[!tp]
\caption{The correctness gate's validation per fault, on runs outside the evaluation campaigns, grouped as in Table~\ref{tab:4}. A check mark means the fault was tested on that model; for fault-free runs, cells give the number of runs. Fault-free runs are replicates, single placement or packing changes, runs of the public recipe and legitimate recipes whose step-0 numerics differ from the baseline. The step-0 loss shift is relative to the model's baseline. Verdict: \emph{pass}, both checks pass; \emph{refused}, the configuration check rejects the candidate before launch; \emph{rejected}, the configuration check cannot see the fault and the loss check rejects it; \emph{missed}, both checks pass a faulty run. We measured the step-0 shifts of refused configurations by running them outside any campaign. In Table~\ref{tab:4}, a fault counts once for each model it was tested on, and each of the 19 uneven pipeline splits proposed in the ablation campaigns counts once; its 29 configuration faults are the 6 argument faults, the 4 stack-defect cases and these 19 splits. The skipped reduction is that of one gradient bucket, the shifted loss mask keeps the token count, and the QK-norm weight gradients are summed instead of averaged over tensor-parallel ranks. $\ddagger$: a real stack defect. $\dagger$: development model. EP, ETP, PP, TP: expert, expert-tensor, pipeline and tensor parallelism.}\label{tab:B2}
\vskip 0.1in
\small
\setlength{\tabcolsep}{2.6pt}
\newcommand{\rh}[2]{\multicolumn{1}{c}{\footnotesize\begin{tabular}[b]{@{}c@{}}#1\\#2\end{tabular}}}
\newcommand{\y}{$\checkmark$}
\begin{tabularx}{\linewidth}{@{}>{\raggedright\arraybackslash}Xcccccccll@{}}
\toprule
Run & \rh{}{GLM-4.5} & \rh{MiniMax}{M2} & \rh{Mixtral}{8x22B} & \rh{Llama-3.1}{405B} & \rh{GPT-OSS}{20B$^\dagger$} & \rh{GPT-OSS}{120B$^\dagger$} & \rh{Qwen3.5}{122B$^\dagger$} & \begin{tabular}[b]{@{}l@{}}Step-0\\loss shift\end{tabular} & Verdict \\
\midrule
\multicolumn{10}{@{}l}{\emph{Legitimate runs}} \\
Fault-free runs & 27 & 31 & 28 & 4 & 32 & 10 & & $\leq$1.98\% & pass \\
\addlinespace[3pt]
\multicolumn{10}{@{}l}{\emph{Configuration faults}} \\
EP = 1 with ETP $\geq$ 4$^\ddagger$ & \y & \y & \y & & & & & $\leq$0.02\% & refused \\
\addlinespace[1.5pt]
PP = 32$^\ddagger$ & \y & & & & & & & 0.00\% & refused \\
\addlinespace[1.5pt]
Capacity factor 1 & & & & & & & \y & +7.6\% & refused \\
\addlinespace[1.5pt]
Learning rate $\times$10 & & & & & & & \y & $-$0.02\% & refused \\
\addlinespace[1.5pt]
Global batch changed & & & & & & & \y & +0.2\% & refused \\
\addlinespace[1.5pt]
BF16 gradient reduction & & & & & & & \y & $-$0.04\% & refused \\
\addlinespace[1.5pt]
Per-token loss off & & & & & & & \y & +0.5\% & refused \\
\addlinespace[1.5pt]
Auxiliary-loss coefficient changed & & & & & & & \y & $-$0.01\% & refused \\
\addlinespace[3pt]
\multicolumn{10}{@{}l}{\emph{Forward-pass code faults}} \\
Labels offset by one token & \y & \y & & \y & & \y & & +190--872\% & rejected \\
\addlinespace[1.5pt]
Attention sinks ignored & & & & & & \y & & +151\% & rejected \\
\addlinespace[3pt]
\multicolumn{10}{@{}l}{\emph{Update-path code faults}} \\
Gradient-bucket reduction skipped & \y & \y & \y & \y & & \y & & 0.00\% & missed \\
\addlinespace[1.5pt]
Load-balancing loss zeroed & \y & & \y & & & & & 0.00\% & missed \\
\addlinespace[1.5pt]
Loss mask shifted by one token & & & & & \y & & & +2.0\% & missed \\
\addlinespace[1.5pt]
QK-norm gradients summed$^\ddagger$ & & \y & & & & & & 0.00\% & missed \\
\bottomrule
\end{tabularx}
\end{table*}

\begin{figure}[!htb]
\centering
\includegraphics[width=\columnwidth]{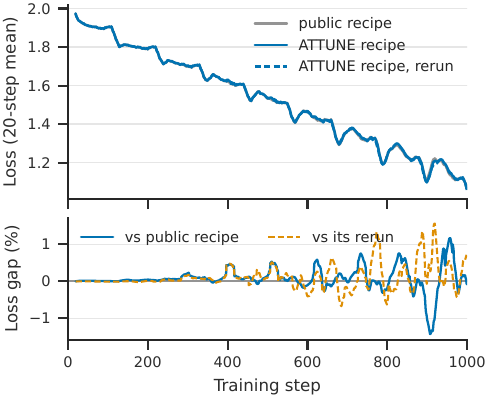}
\caption{MiniMax-M2's public recipe and ATTUNE's shipped recipe from campaign 1, run twice, over 1,000 steps at a constant learning rate of 1e-5. Top: loss as a 20-step moving mean; the three curves overlap. Bottom: the shipped recipe's relative loss gap to the public recipe and to its own rerun, also as 20-step moving means. The gap to the public recipe is about as large as the gap between the shipped recipe's two runs.}\label{fig:longrun}
\end{figure}

\paragraph{Long run.} We trained the shipped recipe of ATTUNE's MiniMax-M2 campaign 1 (Table~\ref{tab:D3}) twice and its public recipe once, for 1,000 steps each (Figure~\ref{fig:longrun}). All three runs use the evaluation workload's rows and seed, at a constant learning rate of 1e-5, ten times the workload's, and make about nine passes over the rows. Against the public recipe, the shipped recipe passes Megatron-Bridge's convergence check over all 1,000 steps, with a correlation of 0.9995. Over steps 900--999 their mean relative loss gap is 0.89\%, against 0.93\% between the two runs of the shipped recipe. Over steps 10--999 the shipped recipe is 2.66 times as fast as the public recipe, close to the 2.71 times of the final re-measurement.

\section{Environment, Versions and Isolation}\label{sec:C}
\setcounter{table}{0}

All runs on a model use one pinned training image per cluster, and each node has 3.2 Tb/s of RDMA networking. Table~\ref{tab:C1} lists the component versions, which are the same in the B200 and H100 images.

\begin{table*}[!tp]
\caption{Software in the pinned training images and the tuners' sandbox.}\label{tab:C1}
\vskip 0.1in
\small
\begin{tabularx}{\linewidth}{@{}l>{\raggedright\arraybackslash}X@{}}
\toprule
Component & Version \\
\midrule
OS & Ubuntu 24.04 \\
Python & 3.12.3 (training image) \\
PyTorch & 2.11.0 (CUDA 12.9 build) \\
CUDA & 12.9 \\
NCCL / libfabric NCCL plugin & 2.28.9 / 1.20.0 \\
Megatron-LM & commit \texttt{1dcf0da} \\
slime & 0.3.0: THUDM/slime \texttt{a897e1f} plus 8 commits that do not change the training step \\
Transformer Engine & 2.10 \\
DeepEP \citep{zhao2025deepep} & 1.2.1 \\
Megatron-Bridge convergence check & commit \texttt{c3c974d}, vendored \\
Claude Code (agentic tuners) & 2.1.286 \\
OptiCo (baseline) & OptiCo-public commit \texttt{e5efc74} \\
Optuna (black-box searches) & 5.0.0 \\
\bottomrule
\end{tabularx}
\end{table*}

\paragraph{Sandbox and broker.} ATTUNE's campaigns receive their skill and a description of the public recipe, including its source. Each campaign runs in a fresh sandbox. The sandbox holds only the campaign's skill, its campaign directory, and the model's correctness rules, checkpoint and data; everything except the campaign directory is read-only. It reaches the cluster only through a \emph{broker}, a proxy that accepts only that campaign's jobs. The broker checks each job specification, applies the stack fixes of Appendix A and logs every request. Through the broker, the agent can also read files inside the training image by running a command in the campaign's own pod. The ablations of Section 3.4 run on miles. In every variant, the same pre-tool hook blocks reads of other campaigns' records.

\paragraph{Isolation audit.} In the 36 evaluation campaigns of Tables~\ref{tab:D2} and~\ref{tab:D3}, the sandbox and broker denied every request for a file or cluster object outside the campaign. No tuner opened any of the ten recipe files for the evaluation models that the training images contain.

\section{Budgets, Measurement and Per-campaign Results}\label{sec:D}
\setcounter{table}{0}

This section gives how campaigns are charged and measured, the statistics behind each comparison, and the result of every evaluation campaign.

\paragraph{Charging.} Every launch is charged, including failed and profiled launches and correctness checks, even when it overshoots the budget. The number of launches is not limited. Every tuner runs at most three jobs at a time. ATTUNE and OptiCo may stop before the budget is spent, while Bayesian optimization and random search run to it.

\paragraph{Budgets and bands.} Table~\ref{tab:D1} gives each model's launch cost, budget and noise band. The band has a 3\% floor.

\begin{table*}[!tp]
\caption{Per-launch cost and budgets, from at least five calibration runs per model on distinct node allocations. The calibration mean is the denominator during search; reported gains use the final re-measurement. $\sigma$ is the calibration runs' relative standard deviation, and each band is max(3\%, 3$\sigma$). The budget is 30 times the unprofiled launch cost, rounded up to a multiple of 10 GPU-hours.}\label{tab:D1}
\vskip 0.1in
\small
\begin{tabular*}{\linewidth}{@{\extracolsep{\fill}}llrrrrr@{}}
\toprule
Model & GPUs & \begin{tabular}[b]{@{}r@{}}Calibration mean\\(tokens/s/GPU)\end{tabular} & $\sigma$ & Band & \begin{tabular}[b]{@{}r@{}}Unprofiled launch\\(GPU-h)\end{tabular} & \begin{tabular}[b]{@{}r@{}}Budget per\\campaign (GPU-h)\end{tabular} \\
\midrule
MiniMax-M2 & 64$\times$B200 & 893.18 & 0.95\% & 3.00\% & 17.58 & 530 \\
GLM-4.5 & 128$\times$B200 & 728.19 & 0.49\% & 3.00\% & 62.58 & 1,880 \\
Llama-3.1-405B & 128$\times$B200 & 399.36 & 1.29\% & 3.88\% & 61.38 & 1,850 \\
Mixtral-8x22B & 128$\times$H100 & 1,091.69 & 0.95\% & 3.00\% & 24.57 & 740 \\
\bottomrule
\end{tabular*}
\end{table*}

\paragraph{Final stage.} After all campaigns on a model, the baseline and the shipped recipes run twice more in one randomized, interleaved order, on the nodes the scheduler assigns. A recipe's value is the mean of all its repeats. Any recipe, the baseline included, got one or two extra repeats when its first two disagreed by more than the band or one ran on a degraded GPU.

\paragraph{Statistics.} Alongside each mean-rule verdict we report an exact one-sided permutation p-value for the difference of means. With three campaigns per tuner, it considers all 20 ways to split the six campaign values between the two tuners; for the five-campaign ablations, all 252 ways to split the ten values.

\paragraph{Per-campaign results.} Tables~\ref{tab:D2} and~\ref{tab:D3} give every evaluation campaign: its shipped throughput and gain, the GPU-hours it used, why it stopped and the changes it shipped.

\begin{table*}[!tp]
\caption{Every evaluation campaign on GLM-4.5, in tokens/s/GPU from the final re-measurement. Gain is over the re-measured public recipe (723.6). GPU-hours include correctness checks. Stop: the agent's convergence judgment, the GPU-hour budget or our OptiCo driver's stop after 10 consecutive iterations that launch nothing (no launch). BO: Bayesian optimization; TPE: tree-structured Parzen estimator. Changes are relative to the public recipe; none: the campaign ships the public recipe. TP, PP, EP, ETP, CP, DP: tensor, pipeline, expert, expert-tensor, context and data parallelism; VP: virtual stages per pipeline rank; first/last-stage layers: transformer layers in the first and last pipeline stages.}\label{tab:D2}
\vskip 0.1in
\small
\begin{tabularx}{\linewidth}{@{}>{\hsize=1.130\hsize\raggedright\arraybackslash}X>{\hsize=0.210\hsize\raggedleft\arraybackslash}X>{\hsize=0.490\hsize\raggedleft\arraybackslash}X>{\hsize=0.490\hsize\raggedleft\arraybackslash}X>{\hsize=0.490\hsize\raggedleft\arraybackslash}X>{\hsize=0.595\hsize\raggedright\arraybackslash}X>{\hsize=3.595\hsize\raggedright\arraybackslash}X@{}}
\toprule
Tuner & \# & Shipped & Gain & GPU-h used & Stop & Shipped changes \\
\midrule
ATTUNE & 1 & 885.7 & +22.4\% & 753.6 & agent & PP 8$\rightarrow$16; EP 16$\rightarrow$8; first/last-stage layers 10/10$\rightarrow$4/4; DeepEP dispatcher; shared-expert overlap off \\
\addlinespace[1.5pt]
 & 2 & 1,119.3 & +54.7\% & 626.3 & agent & TP 2$\rightarrow$1; PP 8$\rightarrow$16; EP 16$\rightarrow$8; VP off$\rightarrow$2; first/last-stage layers 10/10$\rightarrow$4/4; full recompute \\
\addlinespace[1.5pt]
 & 3 & 902.1 & +24.7\% & 627.3 & agent & EP 16$\rightarrow$8; DeepEP dispatcher; shared-expert overlap off; DP overlaps off \\
\addlinespace[5pt]
OptiCo & 1 & 723.6 & 0.0\% & 1,274.4 & no launch & none \\
\addlinespace[1.5pt]
 & 2 & 723.6 & 0.0\% & 1,543.8 & budget & none \\
\addlinespace[1.5pt]
 & 3 & 723.6 & 0.0\% & 1,402.2 & no launch & none \\
\addlinespace[5pt]
BO (TPE) & 1 & 762.8 & +5.4\% & 1,807.8 & budget & PP 8$\rightarrow$4; even stage split; DP overlaps off \\
\addlinespace[1.5pt]
 & 2 & 723.6 & 0.0\% & 1,663.7 & budget & none \\
\addlinespace[1.5pt]
 & 3 & 723.6 & 0.0\% & 1,554.1 & budget & none \\
\addlinespace[5pt]
BO + pre-filter & 1 & 765.6 & +5.8\% & 1,676.7 & budget & PP 8$\rightarrow$4; even stage split; DP overlaps off \\
\addlinespace[1.5pt]
 & 2 & 723.6 & 0.0\% & 1,567.0 & budget & none \\
\addlinespace[1.5pt]
 & 3 & 766.8 & +6.0\% & 1,841.5 & budget & PP 8$\rightarrow$4; even stage split; DP overlaps off \\
\addlinespace[5pt]
Random search & 1 & 723.6 & 0.0\% & 1,546.7 & budget & none \\
\addlinespace[1.5pt]
 & 2 & 892.8 & +23.4\% & 1,787.5 & budget & TP 2$\rightarrow$1; EP 16$\rightarrow$8; CP 1$\rightarrow$2; VP off$\rightarrow$2; no recompute; DP overlaps off \\
\addlinespace[1.5pt]
 & 3 & 723.6 & 0.0\% & 1,729.6 & budget & none \\
\bottomrule
\end{tabularx}
\end{table*}

\begin{table*}[!tp]
\caption{Every evaluation campaign on MiniMax-M2, Llama-3.1-405B and Mixtral-8x22B, as in Table~\ref{tab:D2}. Gain is over the model's re-measured public recipe (MiniMax-M2 895.2, Llama-3.1-405B 394.9, Mixtral-8x22B 1,072.4).}\label{tab:D3}
\vskip 0.1in
\small
\begin{tabularx}{\linewidth}{@{}>{\hsize=1.050\hsize\raggedright\arraybackslash}X>{\hsize=0.210\hsize\raggedleft\arraybackslash}X>{\hsize=0.490\hsize\raggedleft\arraybackslash}X>{\hsize=0.560\hsize\raggedleft\arraybackslash}X>{\hsize=0.490\hsize\raggedleft\arraybackslash}X>{\hsize=0.595\hsize\raggedright\arraybackslash}X>{\hsize=3.605\hsize\raggedright\arraybackslash}X@{}}
\toprule
Tuner & \# & Shipped & Gain & GPU-h used & Stop & Shipped changes \\
\midrule
\multicolumn{7}{l}{\textbf{MiniMax-M2 / 64$\times$B200}} \\
\addlinespace[3pt]
ATTUNE & 1 & 2,423.1 & +170.7\% & 200.2 & agent & TP 2$\rightarrow$1; PP 2$\rightarrow$4; first/last-stage layers even$\rightarrow$15/15; 2 rows per microbatch; DP overlaps on; optimizer offload off \\
\addlinespace[1.5pt]
 & 2 & 2,235.8 & +149.8\% & 197.3 & agent & PP 2$\rightarrow$8; first/last-stage layers even$\rightarrow$7/7; 4 rows per microbatch; optimizer offload off \\
\addlinespace[1.5pt]
 & 3 & 2,436.3 & +172.2\% & 213.0 & agent & TP 2$\rightarrow$1; PP 2$\rightarrow$8; VP off$\rightarrow$2; first/last-stage layers even$\rightarrow$8/6; recompute uniform$\rightarrow$block, 2 layers; optimizer offload off \\
\addlinespace[5pt]
OptiCo & 1 & 1,101.3 & +23.0\% & 301.4 & no launch & TP 2$\rightarrow$1; DP overlaps on; cross-entropy fusion on \\
\addlinespace[1.5pt]
 & 2 & 1,109.9 & +24.0\% & 384.5 & no launch & TP 2$\rightarrow$1; DP overlaps on; cross-entropy fusion on \\
\addlinespace[1.5pt]
 & 3 & 1,077.0 & +20.3\% & 473.0 & budget & 2 rows per microbatch; DP overlaps on; cross-entropy fusion on \\
\addlinespace[5pt]
BO (TPE) & 1 & 1,651.2 & +84.5\% & 462.9 & budget & PP 2$\rightarrow$4; VP off$\rightarrow$2; first/last-stage layers even$\rightarrow$16/14; 2 rows per microbatch; DeepEP dispatcher; optimizer offload off \\
\addlinespace[1.5pt]
 & 2 & 998.5 & +11.5\% & 535.7 & budget & selective recompute of LayerNorm and the MoE activation \\
\addlinespace[1.5pt]
 & 3 & 1,285.9 & +43.6\% & 477.8 & budget & VP off$\rightarrow$31; optimizer offload off \\
\addlinespace[5pt]
BO + pre-filter & 1 & 1,349.4 & +50.7\% & 467.5 & budget & optimizer offload off \\
\addlinespace[1.5pt]
 & 2 & 1,384.1 & +54.6\% & 503.2 & budget & PP 2$\rightarrow$4; first/last-stage layers even$\rightarrow$15/15; optimizer offload off \\
\addlinespace[1.5pt]
 & 3 & 1,441.2 & +61.0\% & 496.9 & budget & TP 2$\rightarrow$1; PP 2$\rightarrow$8; EP 8$\rightarrow$4; CP 1$\rightarrow$2; first/last-stage layers even$\rightarrow$7/7; selective recompute of the MoE layers; DeepEP dispatcher; optimizer offload off \\
\addlinespace[5pt]
Random search & 1 & 961.8 & +7.4\% & 473.7 & budget & PP 2$\rightarrow$4; EP 8$\rightarrow$2; ETP 1$\rightarrow$2; CP 1$\rightarrow$2; first/last-stage layers even$\rightarrow$15/15; 4 rows per microbatch; DeepEP dispatcher; optimizer offload off \\
\addlinespace[1.5pt]
 & 2 & 1,001.2 & +11.8\% & 517.7 & budget & TP 2$\rightarrow$1; PP 2$\rightarrow$16; EP 8$\rightarrow$2; CP 1$\rightarrow$4; VP off$\rightarrow$2; first/last-stage layers even$\rightarrow$2/4; 2 rows per microbatch; optimizer offload off \\
\addlinespace[1.5pt]
 & 3 & 1,599.6 & +78.7\% & 470.6 & budget & TP 2$\rightarrow$1; PP 2$\rightarrow$4; EP 8$\rightarrow$4; first/last-stage layers even$\rightarrow$15/15; optimizer offload off \\
\midrule
\multicolumn{7}{l}{\textbf{Llama-3.1-405B / 128$\times$B200}} \\
\addlinespace[3pt]
ATTUNE & 1 & 438.5 & +11.1\% & 317.8 & agent & PP 8$\rightarrow$16; VP 8$\rightarrow$2; DP 2$\rightarrow$1 \\
\addlinespace[1.5pt]
 & 2 & 415.3 & +5.2\% & 273.8 & agent & VP 8$\rightarrow$2; first/last-stage layers 16/14 instead of counting the embedding and loss as layers \\
\addlinespace[1.5pt]
 & 3 & 411.8 & +4.3\% & 590.0 & agent & VP 8$\rightarrow$4 \\
\midrule
\multicolumn{7}{l}{\textbf{Mixtral-8x22B / 128$\times$H100}} \\
\addlinespace[3pt]
ATTUNE & 1 & 1,072.4 & 0.0\% & 204.8 & agent & none \\
\addlinespace[1.5pt]
 & 2 & 1,072.4 & 0.0\% & 194.1 & agent & none \\
\addlinespace[1.5pt]
 & 3 & 1,072.4 & 0.0\% & 121.4 & agent & none \\
\bottomrule
\end{tabularx}
\end{table*}

\paragraph{Efficiency metrics.} \emph{Wall-clock} runs from a campaign's first cluster admission to its stop record, excluding intervals in which all its jobs wait in the queue. \emph{Launches} counts search launches, excluding correctness checks. \emph{Failed launches} is the share of launches that the control plane labels out of memory, hang or crash. It excludes refused launches, infrastructure failures and runs stopped early by the tuner, its early-abort rule or the harness.

\paragraph{TFLOP/s and MFU.} Table~\ref{tab:A3} converts Mixtral-8x22B's throughput to TFLOP/s with Megatron's FLOP count per token, to compare with the published figure. Tables~\ref{tab:1} and~\ref{tab:A3} report MFU = F $\times$ throughput $\div$ peak, where:

\begin{itemize}
\item \emph{F}, the model FLOPs per token, is 6 $\times$ (dense parameters excluding the input embedding + active expert parameters) plus the attention matrix multiplications, counted with causal masking;
\item \emph{throughput} is tokens/s/GPU;
\item \emph{peak} is the GPU's dense peak at the recipe's precision, as NVIDIA's Exemplar Performance recipes use for MFU \citep{nvidia2026exemplar}: H100 BF16 989 TFLOP/s, B200 BF16 2,250 TFLOP/s and B200 FP8 4,500 TFLOP/s.
\end{itemize}

\paragraph{Payback.} \[\text{payback days} = \frac{\text{campaign GPU-hours}}{\text{GPUs}\times 24\times\bigl(1-\frac{1}{1+\text{gain}}\bigr)},\] with the GPU-hours the campaign actually used. This is the training time after which a run on the tuned recipe has saved the campaign's cost. Payback ignores the agent's LLM cost, about \$58 per MiniMax-M2 campaign and \$112 per GLM-4.5 campaign at list price by Claude Code's accounting. At its mean campaign cost, ATTUNE pays for itself after 0.2 days on MiniMax-M2, 0.9 days on GLM-4.5 and 2.0 days on Llama-3.1-405B. The other tuners take 0.9--1.4 days on MiniMax-M2. On GLM-4.5 random search takes 7.6 days and Bayesian optimization with the pre-filter 14.6 days. On GLM-4.5 payback is omitted for OptiCo, whose campaigns there all ship the public recipe, and for Bayesian optimization without the pre-filter, whose mean gain stays inside the noise band. These figures assume the measured speedup holds over a full training run. On MiniMax-M2 a 1,000-step run kept it (Appendix B).

\paragraph{Reproduced set.} The four NeMo AutoModel campaigns of Table~\ref{tab:1} run NVIDIA's benchmark recipes in its 26.08 container on H100 nodes; GPT-OSS-20B adds upstream fix \#3926 for an attention-bias NaN. They use the same harness, control-plane scripts and correctness gate, through an AutoModel stack adapter. Each run keeps the recipe's run length and scored window: 30 iterations scored over 10--29 on one node, and 20 steps scored over 5--19 for Llama-3.1-70B on eight nodes. The real-text checks train from the released weights on Tulu 3 SFT rows built with the OpenRecipes builder (Appendix~\ref{sec:A}), with the model's own router and a constant learning rate. They compare the shipped recipe with the public recipe under Megatron-Bridge's convergence check. All four pass. Nemotron-3-Nano's full recipe exceeds GPU memory on real text, so its check uses full recomputation in place of selective recomputation, a placement change. Its logged losses match the public recipe's at 58 of 60 steps. On real text, GPT-OSS-20B's public recipe also needs PyTorch's expandable\_segments allocator setting to fit, which both recipes use.

\paragraph{Ablation campaigns.} The 40 campaigns of Table~\ref{tab:ablations} run each job on one 8$\times$H200 node, on miles v0.1.0 \citep{radixark2026miles}. Each model's baseline starts from the single-node GRPO recipe miles publishes for it and trains with SFT instead, on 8,192-token rows of the Tulu 3 SFT mixture \citep{lambert2025tulu} at global batch 128. Web tools are disabled. Each run lasts 20 steps, scored over steps 5--19. Each campaign may spend 20 times its model's mean calibration-run cost, about 45 GPU-hours, and ships by the rule of Section~\ref{sec:2.3}.

\begin{table*}[!tp]
\caption{Ablations of Section~\ref{sec:3.4} (Figure~\ref{fig:ablations}) on GLM-4.7-Flash and Nemotron-3-Nano / 8$\times$H200, five campaigns per variant. Gain is mean $\pm$ standard deviation, each campaign measured against its own run of the public recipe. p is a one-sided exact permutation test of the mean gain against full ATTUNE. Failed launches crashed or did not complete. GPU-h to $\geq$20\% gain is the median GPU-hours a campaign spends before its first gate-passing run with that gain, over the campaigns that reach one (count in parentheses). Bold marks each model's best value.}\label{tab:ablations}
\vskip 0.1in
\small
\begin{tabularx}{\linewidth}{@{}>{\hsize=1.250\hsize\raggedright\arraybackslash}X>{\hsize=1.000\hsize\raggedright\arraybackslash}X>{\hsize=0.667\hsize\raggedleft\arraybackslash}X>{\hsize=0.833\hsize\raggedleft\arraybackslash}X>{\hsize=1.250\hsize\raggedleft\arraybackslash}X@{}}
\toprule
Variant & Gain (\%) & p-value & Failed launches & GPU-h to $\geq$20\% gain \\
\midrule
\addlinespace[4pt]
\multicolumn{5}{l}{\emph{GLM-4.7-Flash}} \\
\addlinespace[4pt]
Full ATTUNE & \textbf{+88.0} $\pm$ 7.1 & -- & 17\% & \textbf{3.9} (5/5) \\
w/o skill & +66.9 $\pm$ 15.8 & 0.016 & 50\% & 9.0 (5/5) \\
w/o open action space & +18.1 $\pm$ 2.7 & 0.004 & \textbf{13\%} & 19.8 (2/5) \\
w/o diagnosis & +73.4 $\pm$ 29.7 & 0.24 & 19\% & 4.9 (5/5) \\
\midrule
\addlinespace[4pt]
\multicolumn{5}{l}{\emph{Nemotron-3-Nano}} \\
\addlinespace[4pt]
Full ATTUNE & \textbf{+49.4} $\pm$ 1.6 & -- & \textbf{32\%} & 5.2 (5/5) \\
w/o skill & +42.9 $\pm$ 5.4 & 0.008 & 50\% & 7.2 (5/5) \\
w/o open action space & +48.6 $\pm$ 4.2 & 0.42 & 33\% & \textbf{3.1} (5/5) \\
w/o diagnosis & +47.5 $\pm$ 3.4 & 0.14 & 41\% & \textbf{3.1} (5/5) \\
\bottomrule
\end{tabularx}
\end{table*}

\section{Baseline Tuners}\label{sec:E}
\setcounter{table}{0}

This section describes how we run OptiCo and the black-box searches on our stack.

\paragraph{OptiCo.} We run the public OptiCo repository at commit \texttt{e5efc74}, with its analyzer and prompts unchanged. We supply three parts of our own: the LLM, the slime adapter and the loop driver. The LLM is the one ATTUNE uses, at its default sampling temperature rather than OptiCo's 0.7. The repository's driver runs 20 cycles by default, while ours runs OptiCo until the budget is spent or 10 consecutive iterations launch nothing.

Our adapter connects OptiCo to slime. It takes the sequence length, global batch and data from the model's frozen baseline, and maps OptiCo's layer-count knobs onto Megatron's pipeline split. OptiCo keeps its 11-knob schema unchanged, and the adapter maps each proposed key and value or rejects it before launch. It rejects the microbatch knob where a model freezes one row per microbatch, and any change the model's frozen arguments forbid. It also applies each model's required co-changes, such as turning the data-parallel overlaps off on GLM-4.5 when expert data parallelism exceeds 1, so that a schema-valid proposal maps to a valid configuration where one exists. Proposals the adapter still rejects do not count toward the driver's stop. OptiCo uses the same control plane, correctness gate and broker as the other tuners. A proposal identical to a configuration the campaign already ran is answered with that run's measurement at no cost and counts as a cached iteration.

\paragraph{Black-box search.} The searches run on MiniMax-M2 and GLM-4.5, each over its frozen space (Table~\ref{tab:E1}). The same space bounds the memory pre-filter's draws.

The searched knobs change placement or are same-math swaps, such as the dispatcher and, on MiniMax-M2, the optimizer-offload toggle, which the baseline sets on. On GLM-4.5 no tuner can reach optimizer offload: Megatron allows it only with the precision-aware optimizer, a numeric setting that the configuration check freezes and that GLM-4.5's public recipe does not use. The space keeps cross-entropy fusion at the baseline's value. On GLM-4.5, the data-parallel overlaps and shared-expert overlap follow the same required co-changes as in OptiCo's adapter: the data-parallel overlaps turn off when expert data parallelism exceeds 1, and shared-expert overlap runs only with the all-to-all dispatcher. A draw that violates a divisibility or layer-count constraint read from the framework source, or hits a known stack defect (Table~\ref{tab:B1}), scores zero without launching.

\begin{table*}[!tp]
\caption{Search spaces of the baseline tuners. A value spanning both columns applies to both models. On both models DeepEP is offered only within a node, and draws across nodes are infeasible and score zero. On GLM-4.5, where DeepEP cannot initialize across nodes on our network, it also runs with shared-expert overlap off. The largest virtual-stage values put one layer in each virtual stage, and the light splits give the first, last or both end stages fewer layers. $^*$: expert parallelism 1 only with expert-tensor parallelism $\leq$ 2.}\label{tab:E1}
\vskip 0.1in
\small
\begin{tabular*}{\linewidth}{@{\extracolsep{\fill}}lll@{}}
\toprule
Knob & MiniMax-M2 (64 GPUs) & GLM-4.5 (128 GPUs) \\
\midrule
Tensor parallelism & \multicolumn{2}{l}{1, 2, 4, 8} \\
Pipeline parallelism & \multicolumn{2}{l}{1, 2, 4, 8, 16} \\
Expert parallelism & 1$^*$, 2, 4, 8, 16, 32, 64 & 1$^*$, 2, 4, 8, 16, 32 \\
Expert-tensor parallelism & \multicolumn{2}{l}{1, 2, 4, 8} \\
Virtual stages per rank & off, 2, 31 & off, 2, 4, 23 \\
Pipeline split & \multicolumn{2}{l}{balanced, light first, light last, light both} \\
Context parallelism & 1, 2, 4, 8 & 1, 2, 4 \\
Rows per microbatch & 1, 2, 4, 32 & 1 (frozen) \\
Recompute & \multicolumn{2}{l}{none; selective (four module sets); full with 1 or 2 layers per checkpoint} \\
Dispatcher & all-to-all; DeepEP within a node & all-to-all; DeepEP within a node \\
Optimizer offload & on (baseline) or off & off (unreachable) \\
\bottomrule
\end{tabular*}
\end{table*}

\paragraph{Settings.} Bayesian optimization uses TPE with 8 startup trials, multivariate sampling and \texttt{constant\_\allowbreak{}liar} for three concurrent trials, with seeds 1--3. Random search samples uniformly, with seeds 1--3. Each search takes the baseline as its first trial. A failure of any kind scores zero: an infeasible draw, a pre-filter rejection, an out-of-memory error, a crash or a hang.

\paragraph{Early abort.} Bayesian optimization and random search stop (prune) a run whose throughput falls below 0.90$\times$ the leader's after at least four steady measured steps. A pruned trial gets no throughput score. Only these two searches use this rule.

\paragraph{Memory pre-filter.} An analytical estimate of peak memory per GPU rejects a draw when the estimate $\times$ (1 + margin) exceeds the GPU's memory. Each model's margin is the largest value that still admits every configuration that fit in memory during the model's bring-up launches: 11.7\% on MiniMax-M2 and 5.8\% on GLM-4.5.

\section{Agent Trace: GLM-4.5 ATTUNE Campaign 2}\label{sec:F}
\setcounter{table}{0}

Table~\ref{tab:F1} lists the campaign's launches in order, with the agent's diagnosis after each run. From the first run's low TFLOP/s and negligible data wait, the agent judged that communication and kernel launches bound the step. Removing tensor parallelism would cut that communication, but it ran out of memory at both PP 8 and PP 16. Adding full recomputation made it fit and raised throughput by 47\%. The agent then tested the leader's neighbors: PP 8 again, an uneven stage split, the DeepEP dispatcher, two virtual stages and CP 2. None beat the leader by more than the band, so the agent stopped, and the fastest candidate, with two virtual stages, shipped.

\begin{table*}[!tp]
\caption{The launches of GLM-4.5 ATTUNE campaign 2 in order, starting from the public recipe (728.2 tokens/s/GPU calibrated). Run names are the agent's labels. Results are in tokens/s/GPU; gains up to and including the winner (\texttt{tp1pp16fr}) are over the public recipe, and later gains are over the winner. Values marked $\approx$ are the agent's estimates for runs it stopped early; the control plane does not score these runs. Each quote is the agent's reasoning about the run; in the quotes, SP is sequence parallelism and a2a is all-to-all. The shipped recipe re-measured at 1,119.3 tokens/s/GPU in the final stage.}\label{tab:F1}
\vskip 0.1in
\small
\begin{tabularx}{\linewidth}{@{}l>{\raggedright\arraybackslash}p{1.95in}>{\raggedright\arraybackslash}p{1.25in}>{\raggedright\arraybackslash}X@{}}
\toprule
Run & Change & Result & The agent's reasoning, quoted \\
\midrule
\texttt{pp16ep8} & PP 8$\rightarrow$16, EP 16$\rightarrow$8, first/last-stage layers 10/10$\rightarrow$4/4, expert all-to-all inside a node & 707.6, tie & ``clearly comm/overhead-bound''; ``EP placement is throughput-neutral here'' \\
\addlinespace[3pt]
\texttt{tp1pp8} & TP 2$\rightarrow$1 & out of memory & ``TP1 (no SP, unsharded activations) doesn't fit at PP8; it needs more PP.'' \\
\addlinespace[3pt]
\texttt{tp1pp16} & TP 2$\rightarrow$1 at PP 16, EP 8 & out of memory & ``memory-infeasible at both PP8 and PP16 with selective recompute'' \\
\addlinespace[3pt]
\texttt{tp1pp16fr} & + full recomputation & 1,072.7, win (+47.3\%) & ``our step has many tiny 1-row microbatches $\rightarrow$ lots of launch idle \ldots{} full recompute may be \textasciitilde{}free here AND it enables the TP1 fit'' \\
\addlinespace[3pt]
\texttt{tp1pp8fr} & back to PP 8, EP 16 & stopped early, $\approx$668 & ``the cross-node EP a2a (and DP16 comm) is now exposed'' \\
\addlinespace[3pt]
\texttt{tp1pp16s53} & first/last-stage layers 4/4$\rightarrow$5/3 & stopped early, $\approx$1,083 & ``the pipeline is not stage-imbalance-bound; it's uniformly comm/launch-bound'' \\
\addlinespace[3pt]
\texttt{tp1pp16dep} & DeepEP dispatcher & stopped early, $\approx$1,022 & ``NCCL's intranode alltoall already beats DeepEP's kernels (which also reserve SMs)'' \\
\addlinespace[3pt]
\texttt{tp1pp16vpp2} & two virtual stages & 1,090.3, tie (+1.6\%), fastest & ``extra per-chunk launches offset the gain on this launch-bound workload'' \\
\addlinespace[3pt]
\texttt{tp1pp16cp2} & CP 1$\rightarrow$2 & stopped early, $\approx$602 & ``decisive lose, closing the CP axis \ldots{} All structural neighbors are now measured.'' \\
\addlinespace[3pt]
\texttt{check} & check run of \texttt{tp1pp16vpp2} & passes both gate checks; ships & -- \\
\bottomrule
\end{tabularx}
\end{table*}

\section{Development Models}\label{sec:G}
\setcounter{table}{0}

\begin{table*}[!tp]
\caption{The catalog audit. The first check compares the 155 lessons that the development campaigns distilled into the catalog with those campaigns' own records: a lesson is wrong when the records or the current framework source contradict it and unsupported when no record supports it. The second asks the current model (Claude Opus 5.5, without the skill, three answers per question) about each of the 106 lesson and policy entries of the curated catalog file, which do not map one-to-one to the 155 checked lessons; an entry is answered partly when the answers miss a condition, scope or failure mode it states.}\label{tab:G1}
\vskip 0.1in
\small
\begin{tabular*}{\linewidth}{@{\extracolsep{\fill}}llr@{}}
\toprule
Check & Outcome & Entries \\
\midrule
Lessons against the campaigns' own records & correct and general & 71 \\
 & rewritten to their general form & 34 \\
 & wrong & 31 \\
 & unsupported & 11 \\
 & rests on one model or campaign & 8 \\
\addlinespace[4pt]
Current model asked without the skill & answered fully & 19 \\
 & answered partly & 59 \\
 & contradicted & 27 \\
 & not known & 1 \\
\bottomrule
\end{tabular*}
\end{table*}

ATTUNE was developed on tuning campaigns over Qwen3.5-122B-A10B, Qwen3.5-397B-A17B and GLM-5.2-744B-A40B across H100, H200 and B200, and over Qwen3.5-35B-A3B, GPT-OSS-20B and GPT-OSS-120B on H100. Its development also used ablation testbeds over Qwen3-235B-A22B, Qwen3-30B-A3B, Qwen3-MoE-A0.6B and GPT3-7B.

During those campaigns the agent wrote lessons back to the lever catalog. The frozen catalog states them as general rules, with no per-campaign numbers or summaries. No held-out model contributed to it.

\paragraph{Catalog audit.} Table~\ref{tab:G1} summarizes two checks of the catalog made after the evaluation (Section 6.1).

\end{document}